%% file: BPH-10-002_temp.tex
\pdfoutput=1
\documentclass[11pt,twoside,a4paper,cmspaper,final,collab]{cms-tdr}
\def\svnVersion{22107:22136M}\def\cmsCernNo{2010-046}\def\cmsCernDate{2010/11/18}\def\cmsMessage{Submitted to the European Physical Journal C}

\begin{document}\cmsNoteHeader{BPH-10-002}
%%%%%%%%%%%%%%%%%%%%%%%%%%%%%%%%%%%%%%%%%%%%%%%%%%%%%%%%%%%%%%%%%%%%
%
%  Common definitions
%
%  N.B. use of \providecommand rather than \newcommand means
%       that a definition is ignored if already specified
%
%                                              L. Taylor 18 Feb 2005
%%%%%%%%%%%%%%%%%%%%%%%%%%%%%%%%%%%%%%%%%%%%%%%%%%%%%%%%%%%%%%%%%%%%

%%%%%%%%%%%%%%%%%%%%%%%%%%%%%%%%%%%%%%%%%%%%%%%%%%%%%%%%%%%%%%%%%%%%
%
% Hyphenations (only need to add here if you get a nasty word break)
%
\hyphenation{env-iron-men-tal}%    just an example
\hyphenation{had-ron-i-za-tion}
\hyphenation{cal-or-i-me-ter}
\hyphenation{de-vices}
%
% Hyphenations-end
%\% svn info. These are modified by svn at checkout time.
% The last version of these macros found before the maketitle will be the one on the front page,
% so only the main file is tracked.
% Do not edit by hand!
\RCS$Revision: 22109 $
\RCS$HeadURL: svn+ssh://alverson@svn.cern.ch/reps/tdr2/papers/BPH-10-002/trunk/BPH-10-002.tex $
\RCS$Id: BPH-10-002.tex 22109 2010-11-18 01:40:27Z alverson $
%%%%%%%%%%%%% ptdr definitions %%%%%%%%%%%%%%%%%%%%%
\input{ptdr-definitions}
\newcommand{\jpsi}{{\ensuremath{J/\psi}}}
\newcommand{\ups}{$\Upsilon$ }
\newcommand{\cc}{$\mbox{c\bar{c}}$ }
\newcommand{\ptjpsi}{{\ensuremath{p_{\mathrm{T}}^{J/\psi}}}}
\newcommand{\bb}{$b\bar b$}
\newcommand{\etajpsi}{{\ensuremath{y}}}
\newcommand{\ptmu}{{\ensuremath{p_{\mathrm{T}}^{\mu}}}}
\newcommand{\etamu}{{\ensuremath{\eta^{\mu}}}}
\newcommand{\dsigmadpt}{{\ensuremath{d^2\sigma/(dp_\mathrm{T} dy)}}}
\newcommand{\ptb}{{\ensuremath{p_T^{H_b}}}}
\newcommand{\effic}{{\ensuremath{\epsilon}}}

%%%%%%%%%%%%%%%  Title page %%%%%%%%%%%%%%%%%%%%%%%%
\cmsNoteHeader{BPH-10-002} % This is over-written in the CMS environment: useful as preprint no. for export versions
\title{Prompt and non-prompt $J/\psi$ production in pp collisions at $\sqrt{s} = 7$~TeV}% Force line breaks with \\

%Author is always "The CMS Collaboration" for PAS and papers, so author, etc, below will be ignored in those cases
\address[QTF]{CERN}
\author[QTF]{The JPsi team}

% please supply the date in yyyy/mm/dd format. Today has been
% redefined to do so, but it should be fixed as of the final release date.
% For papers and PAS, \today is taken as the date the head file (this one) was last modified according to svn: see the RCS Id string above.
\date{\today}

% Abstract processing:
% 1. **DO NOT use \include or \input** to include the abstract: our abstract extractor will not search through other files than this one.
% 2. **DO NOT use %** to comment out sections of the abstract: the extractor will still grab those lines (and they won't be comments any longer!).
% 3. **DO NOT use tex macros** in the abstract: External TeX parsers used on the abstract don't understand them.
\abstract{
The production of $J/\psi$ mesons is studied in pp collisions at $\sqrt s = 7$ TeV with the CMS experiment at the LHC.
The measurement is based on a dimuon sample corresponding to an integrated luminosity of 314~nb$^{-1}$.
The $J/\psi$ differential cross section is determined, as a function of the $J/\psi$
transverse momentum, in three rapidity ranges.  A fit to the decay length distribution is used to separate the prompt from the non-prompt (b hadron to $J/\psi$) component.
Integrated over $J/\psi$ transverse momentum from 6.5 to 30~GeV/$c$
and over rapidity in the range $|y| < 2.4$, the measured cross
sections, times the dimuon decay branching fraction, are $70.9 \pm
2.1 (\mbox{stat.}) \pm 3.0 (\mbox{syst.}) \pm 7.8
(\mbox{luminosity})$~nb for prompt $J/\psi$ mesons
assuming unpolarized production and $26.0 \pm
1.4 (\mbox{stat.}) \pm 1.6 (\mbox{syst.}) \pm 2.9 (\mbox{luminosity})$~nb
for $J/\psi$ mesons from b-hadron decays.
}

% Do not comment out the following hypersetup lines (metadata). They will disappear in NODRAFT mode and are needed by CDS.
% Also: make sure that the values of the metadata items are sensible.
\hypersetup{%
pdfauthor={CMS Collaboration},%
pdftitle={Prompt and non-prompt J/psi production in pp collisions at sqrt(s) = 7 TeV},%
pdfsubject={CMS},%
pdfkeywords={CMS, physics, JPsi, cross section, Quarkonia}}

\maketitle %maketitle comes after all the front information has been supplied

%%%%%%%%%%%%%%%%%%%%%%%%%%%%%%%%  Begin text %%%%%%%%%%%%%%%%%%%%%%%%%%%%%

\newboolean{DEBUG}
%\setboolean{DEBUG}{true}
\setboolean{DEBUG}{false}

\input{Introduction}
\input{CMS}
\input{DataSelection}

\input{Acceptance_efficiency}

\input{Inclusive}

\input{Bfraction}
\input{Comparison}
\input{Conclusions}

\newpage
\clearpage

\bibliography{auto_generated}   % will be created by the tdr script.
\clearpage

\cleardoublepage\appendix\section{The CMS Collaboration \label{app:collab}}\begin{sloppypar}\hyphenpenalty=5000\widowpenalty=500\clubpenalty=5000\input{BPH-10-002-authorlist.tex}\end{sloppypar}
\end{document}

%% file: ptdr-definitions.tex
%%%%%%%%%%%%%%%%%%%%%%%%%%%%%%%%%%%%%%%%%%%%%%%%%%%%%%%%%%%%%%%%%%%%
%
%  Common definitions
%
%  N.B. use of \providecommand rather than \newcommand means
%       that a definition is ignored if already specified
%
%                                              L. Taylor 18 Feb 2005
%%%%%%%%%%%%%%%%%%%%%%%%%%%%%%%%%%%%%%%%%%%%%%%%%%%%%%%%%%%%%%%%%%%%

% Some shorthand
% turn off italics
\providecommand {\etal}{\mbox{et al.}\xspace} %et al. - no preceding comma
\providecommand {\ie}{\mbox{i.e.}\xspace}     %i.e.
\providecommand {\eg}{\mbox{e.g.}\xspace}     %e.g.
\providecommand {\etc}{\mbox{etc.}\xspace}     %etc.
\providecommand {\vs}{\mbox{\sl vs.}\xspace}      %vs.
\providecommand {\mdash}{\ensuremath{\mathrm{-}}} % for use within formulas

% some terms whose definition we may change
\providecommand {\Lone}{Level-1\xspace} % Level-1 or L1 ?
\providecommand {\Ltwo}{Level-2\xspace}
\providecommand {\Lthree}{Level-3\xspace}

% Some software programs (alphabetized)
\providecommand{\ACERMC} {\textsc{AcerMC}\xspace}
\providecommand{\ALPGEN} {{\textsc{alpgen}}\xspace}
\providecommand{\CHARYBDIS} {{\textsc{charybdis}}\xspace}
\providecommand{\CMKIN} {\textsc{cmkin}\xspace}
\providecommand{\CMSIM} {{\textsc{cmsim}}\xspace}
\providecommand{\CMSSW} {{\textsc{cmssw}}\xspace}
\providecommand{\COBRA} {{\textsc{cobra}}\xspace}
\providecommand{\COCOA} {{\textsc{cocoa}}\xspace}
\providecommand{\COMPHEP} {\textsc{CompHEP}\xspace}
\providecommand{\EVTGEN} {{\textsc{evtgen}}\xspace}
\providecommand{\FAMOS} {{\textsc{famos}}\xspace}
\providecommand{\GARCON} {\textsc{garcon}\xspace}
\providecommand{\GARFIELD} {{\textsc{garfield}}\xspace}
\providecommand{\GEANE} {{\textsc{geane}}\xspace}
\providecommand{\GEANTfour} {{\textsc{geant4}}\xspace}
\providecommand{\GEANTthree} {{\textsc{geant3}}\xspace}
\providecommand{\GEANT} {{\textsc{geant}}\xspace}
\providecommand{\HDECAY} {\textsc{hdecay}\xspace}
\providecommand{\HERWIG} {{\textsc{herwig}}\xspace}
\providecommand{\HIGLU} {{\textsc{higlu}}\xspace}
\providecommand{\HIJING} {{\textsc{hijing}}\xspace}
\providecommand{\IGUANA} {\textsc{iguana}\xspace}
\providecommand{\ISAJET} {{\textsc{isajet}}\xspace}
\providecommand{\ISAPYTHIA} {{\textsc{isapythia}}\xspace}
\providecommand{\ISASUGRA} {{\textsc{isasugra}}\xspace}
\providecommand{\ISASUSY} {{\textsc{isasusy}}\xspace}
\providecommand{\ISAWIG} {{\textsc{isawig}}\xspace}
\providecommand{\MADGRAPH} {\textsc{MadGraph}\xspace}
\providecommand{\MCATNLO} {\textsc{mc@nlo}\xspace}
\providecommand{\MCFM} {\textsc{mcfm}\xspace}
\providecommand{\MILLEPEDE} {{\textsc{millepede}}\xspace}
\providecommand{\ORCA} {{\textsc{orca}}\xspace}
\providecommand{\OSCAR} {{\textsc{oscar}}\xspace}
\providecommand{\PHOTOS} {\textsc{photos}\xspace}
\providecommand{\PROSPINO} {\textsc{prospino}\xspace}
\providecommand{\PYTHIA} {{\textsc{pythia}}\xspace}
\providecommand{\SHERPA} {{\textsc{sherpa}}\xspace}
\providecommand{\TAUOLA} {\textsc{tauola}\xspace}
\providecommand{\TOPREX} {\textsc{TopReX}\xspace}
\providecommand{\XDAQ} {{\textsc{xdaq}}\xspace}

%  Experiments
\providecommand {\DZERO}{D\O\xspace}     %etc.

% Measurements and units...

\providecommand{\de}{\ensuremath{^\circ}}
\providecommand{\ten}[1]{\ensuremath{\times \text{10}^\text{#1}}}
\providecommand{\unit}[1]{\ensuremath{\text{\,#1}}\xspace}
\providecommand{\mum}{\ensuremath{\,\mu\text{m}}\xspace}
\providecommand{\micron}{\ensuremath{\,\mu\text{m}}\xspace}
\providecommand{\cm}{\ensuremath{\,\text{cm}}\xspace}
\providecommand{\mm}{\ensuremath{\,\text{mm}}\xspace}
\providecommand{\mus}{\ensuremath{\,\mu\text{s}}\xspace}
\providecommand{\keV}{\ensuremath{\,\text{ke\hspace{-.08em}V}}\xspace}
\providecommand{\MeV}{\ensuremath{\,\text{Me\hspace{-.08em}V}}\xspace}
\providecommand{\GeV}{\ensuremath{\,\text{Ge\hspace{-.08em}V}}\xspace}
\providecommand{\TeV}{\ensuremath{\,\text{Te\hspace{-.08em}V}}\xspace}
\providecommand{\PeV}{\ensuremath{\,\text{Pe\hspace{-.08em}V}}\xspace}
\providecommand{\keVc}{\ensuremath{{\,\text{ke\hspace{-.08em}V\hspace{-0.16em}/\hspace{-0.08em}}c}}\xspace}
\providecommand{\MeVc}{\ensuremath{{\,\text{Me\hspace{-.08em}V\hspace{-0.16em}/\hspace{-0.08em}}c}}\xspace}
\providecommand{\GeVc}{\ensuremath{{\,\text{Ge\hspace{-.08em}V\hspace{-0.16em}/\hspace{-0.08em}}c}}\xspace}
\providecommand{\TeVc}{\ensuremath{{\,\text{Te\hspace{-.08em}V\hspace{-0.16em}/\hspace{-0.08em}}c}}\xspace}
\providecommand{\keVcc}{\ensuremath{{\,\text{ke\hspace{-.08em}V\hspace{-0.16em}/\hspace{-0.08em}}c^\text{2}}}\xspace}
\providecommand{\MeVcc}{\ensuremath{{\,\text{Me\hspace{-.08em}V\hspace{-0.16em}/\hspace{-0.08em}}c^\text{2}}}\xspace}
\providecommand{\GeVcc}{\ensuremath{{\,\text{Ge\hspace{-.08em}V\hspace{-0.16em}/\hspace{-0.08em}}c^\text{2}}}\xspace}
\providecommand{\TeVcc}{\ensuremath{{\,\text{Te\hspace{-.08em}V\hspace{-0.16em}/\hspace{-0.08em}}c^\text{2}}}\xspace}

\providecommand{\pbinv} {\mbox{\ensuremath{\,\text{pb}^\text{$-$1}}}\xspace}
\providecommand{\fbinv} {\mbox{\ensuremath{\,\text{fb}^\text{$-$1}}}\xspace}
\providecommand{\nbinv} {\mbox{\ensuremath{\,\text{nb}^\text{$-$1}}}\xspace}
\providecommand{\percms}{\ensuremath{\,\text{cm}^\text{$-$2}\,\text{s}^\text{$-$1}}\xspace}
\providecommand{\lumi}{\ensuremath{\mathcal{L}}\xspace}
\providecommand{\Lumi}{\ensuremath{\mathcal{L}}\xspace}%both upper and lower
%
% Need a convention here:
\providecommand{\LvLow}  {\ensuremath{\mathcal{L}=\text{10}^\text{32}\,\text{cm}^\text{$-$2}\,\text{s}^\text{$-$1}}\xspace}
\providecommand{\LLow}   {\ensuremath{\mathcal{L}=\text{10}^\text{33}\,\text{cm}^\text{$-$2}\,\text{s}^\text{$-$1}}\xspace}
\providecommand{\lowlumi}{\ensuremath{\mathcal{L}=\text{2}\times \text{10}^\text{33}\,\text{cm}^\text{$-$2}\,\text{s}^\text{$-$1}}\xspace}
\providecommand{\LMed}   {\ensuremath{\mathcal{L}=\text{2}\times \text{10}^\text{33}\,\text{cm}^\text{$-$2}\,\text{s}^\text{$-$1}}\xspace}
\providecommand{\LHigh}  {\ensuremath{\mathcal{L}=\text{10}^\text{34}\,\text{cm}^\text{$-$2}\,\text{s}^\text{$-$1}}\xspace}
\providecommand{\hilumi} {\ensuremath{\mathcal{L}=\text{10}^\text{34}\,\text{cm}^\text{$-$2}\,\text{s}^\text{$-$1}}\xspace}

% Some usual physics terms

\providecommand{\zp}{\ensuremath{\mathrm{Z}^\prime}\xspace}

% SM (still to be classified)

\providecommand{\kt}{\ensuremath{k_{\mathrm{T}}}\xspace}
\providecommand{\BC}{\ensuremath{\mathrm{B_{c}}}\xspace}
\providecommand{\bbarc}{\ensuremath{\mathrm{\overline{b}c}}\xspace}
\providecommand{\bbbar}{\ensuremath{\mathrm{b\overline{b}}}\xspace}
\providecommand{\ccbar}{\ensuremath{\mathrm{c\overline{c}}}\xspace}
\providecommand{\JPsi}{\ensuremath{\mathrm{J}\hspace{-.08em}/\hspace{-.14em}\psi}\xspace}
\providecommand{\bspsiphi}{\ensuremath{\mathrm{B_s} \to \JPsi\, \phi}\xspace}
\providecommand{\AFB}{\ensuremath{A_\text{FB}}\xspace}
\providecommand{\EE}{\ensuremath{\mathrm{e^+e^-}}\xspace}
\providecommand{\MM}{\ensuremath{\mu^+\mu^-}\xspace}
\providecommand{\TT}{\ensuremath{\tau^+\tau^-}\xspace}
\providecommand{\wangle}{\ensuremath{\sin^{2}\theta_{\text{eff}}^\text{lept}(M^2_\mathrm{Z})}\xspace}
\providecommand{\ttbar}{\ensuremath{\mathrm{t\overline{t}}}\xspace}
\providecommand{\stat}{\ensuremath{\,\text{(stat.)}}\xspace}
\providecommand{\syst}{\ensuremath{\,\text{(syst.)}}\xspace}

%%%  E-gamma definitions
\providecommand{\HGG}{\ensuremath{\mathrm{H}\to\gamma\gamma}}
\providecommand{\gev}{\GeV}
\providecommand{\GAMJET}{\ensuremath{\gamma + \text{jet}}}
\providecommand{\PPTOJETS}{\ensuremath{\mathrm{pp}\to\text{jets}}}
\providecommand{\PPTOGG}{\ensuremath{\mathrm{pp}\to\gamma\gamma}}
\providecommand{\PPTOGAMJET}{\ensuremath{\mathrm{pp}\to\gamma + \mathrm{jet}}}
\providecommand{\MH}{\ensuremath{M_{\mathrm{H}}}}
\providecommand{\RNINE}{\ensuremath{R_\mathrm{9}}}
\providecommand{\DR}{\ensuremath{\Delta R}}

% Physics symbols ...

\providecommand{\PT}{\ensuremath{p_{\mathrm{T}}}\xspace}
\providecommand{\pt}{\ensuremath{p_{\mathrm{T}}}\xspace}
\providecommand{\ET}{\ensuremath{E_{\mathrm{T}}}\xspace}
\providecommand{\HT}{\ensuremath{H_{\mathrm{T}}}\xspace}
\providecommand{\et}{\ensuremath{E_{\mathrm{T}}}\xspace}
\providecommand{\Em}{\ensuremath{E\hspace{-0.6em}/}\xspace}
\providecommand{\Pm}{\ensuremath{p\hspace{-0.5em}/}\xspace}
\providecommand{\PTm}{\ensuremath{{p}_\mathrm{T}\hspace{-1.02em}/}\xspace}
\providecommand{\PTslash}{\ensuremath{{p}_\mathrm{T}\hspace{-1.02em}/}\xspace}
\providecommand{\ETm}{\ensuremath{E_{\mathrm{T}}^{\text{miss}}}\xspace}
\providecommand{\ETslash}{\ensuremath{E_{\mathrm{T}}\hspace{-1.1em}/}\xspace}
\providecommand{\MET}{\ensuremath{E_{\mathrm{T}}^{\text{miss}}}\xspace}
\providecommand{\ETmiss}{\ensuremath{E_{\mathrm{T}}^{\text{miss}}}\xspace}
\providecommand{\VEtmiss}{\ensuremath{{\vec E}_{\mathrm{T}}^{\text{miss}}}\xspace}

% roman face derivative
\providecommand{\dd}[2]{\ensuremath{\frac{\mathrm{d} #1}{\mathrm{d} #2}}}

%%%%%%
% From Albert
%

\providecommand{\ga}{\ensuremath{\gtrsim}}
\providecommand{\la}{\ensuremath{\lesssim}}
\providecommand{\swsq}{\ensuremath{\sin^2\theta_\mathrm{W}}\xspace}
\providecommand{\cwsq}{\ensuremath{\cos^2\theta_\mathrm{W}}\xspace}
\providecommand{\tanb}{\ensuremath{\tan\beta}\xspace}
\providecommand{\tanbsq}{\ensuremath{\tan^{2}\beta}\xspace}
\providecommand{\sidb}{\ensuremath{\sin 2\beta}\xspace}
\providecommand{\alpS}{\ensuremath{\alpha_S}\xspace}
\providecommand{\alpt}{\ensuremath{\tilde{\alpha}}\xspace}

\providecommand{\QL}{\ensuremath{\mathrm{Q}_\mathrm{L}}\xspace}
\providecommand{\sQ}{\ensuremath{\tilde{\mathrm{Q}}}\xspace}
\providecommand{\sQL}{\ensuremath{\tilde{\mathrm{Q}}_\mathrm{L}}\xspace}
\providecommand{\ULC}{\ensuremath{\mathrm{U}_\mathrm{L}^\mathrm{C}}\xspace}
\providecommand{\sUC}{\ensuremath{\tilde{\mathrm{U}}^\mathrm{C}}\xspace}
\providecommand{\sULC}{\ensuremath{\tilde{\mathrm{U}}_\mathrm{L}^\mathrm{C}}\xspace}
\providecommand{\DLC}{\ensuremath{\mathrm{D}_\mathrm{L}^\mathrm{C}}\xspace}
\providecommand{\sDC}{\ensuremath{\tilde{\mathrm{D}}^\mathrm{C}}\xspace}
\providecommand{\sDLC}{\ensuremath{\tilde{\mathrm{D}}_\mathrm{L}^\mathrm{C}}\xspace}
\providecommand{\LL}{\ensuremath{\mathrm{L}_\mathrm{L}}\xspace}
\providecommand{\sL}{\ensuremath{\tilde{\mathrm{L}}}\xspace}
\providecommand{\sLL}{\ensuremath{\tilde{\mathrm{L}}_\mathrm{L}}\xspace}
\providecommand{\ELC}{\ensuremath{\mathrm{E}_\mathrm{L}^\mathrm{C}}\xspace}
\providecommand{\sEC}{\ensuremath{\tilde{\mathrm{E}}^\mathrm{C}}\xspace}
\providecommand{\sELC}{\ensuremath{\tilde{\mathrm{E}}_\mathrm{L}^\mathrm{C}}\xspace}
\providecommand{\sEL}{\ensuremath{\tilde{\mathrm{E}}_\mathrm{L}}\xspace}
\providecommand{\sER}{\ensuremath{\tilde{\mathrm{E}}_\mathrm{R}}\xspace}
\providecommand{\sFer}{\ensuremath{\tilde{\mathrm{f}}}\xspace}
\providecommand{\sQua}{\ensuremath{\tilde{\mathrm{q}}}\xspace}
\providecommand{\sUp}{\ensuremath{\tilde{\mathrm{u}}}\xspace}
\providecommand{\suL}{\ensuremath{\tilde{\mathrm{u}}_\mathrm{L}}\xspace}
\providecommand{\suR}{\ensuremath{\tilde{\mathrm{u}}_\mathrm{R}}\xspace}
\providecommand{\sDw}{\ensuremath{\tilde{\mathrm{d}}}\xspace}
\providecommand{\sdL}{\ensuremath{\tilde{\mathrm{d}}_\mathrm{L}}\xspace}
\providecommand{\sdR}{\ensuremath{\tilde{\mathrm{d}}_\mathrm{R}}\xspace}
\providecommand{\sTop}{\ensuremath{\tilde{\mathrm{t}}}\xspace}
\providecommand{\stL}{\ensuremath{\tilde{\mathrm{t}}_\mathrm{L}}\xspace}
\providecommand{\stR}{\ensuremath{\tilde{\mathrm{t}}_\mathrm{R}}\xspace}
\providecommand{\stone}{\ensuremath{\tilde{\mathrm{t}}_1}\xspace}
\providecommand{\sttwo}{\ensuremath{\tilde{\mathrm{t}}_2}\xspace}
\providecommand{\sBot}{\ensuremath{\tilde{\mathrm{b}}}\xspace}
\providecommand{\sbL}{\ensuremath{\tilde{\mathrm{b}}_\mathrm{L}}\xspace}
\providecommand{\sbR}{\ensuremath{\tilde{\mathrm{b}}_\mathrm{R}}\xspace}
\providecommand{\sbone}{\ensuremath{\tilde{\mathrm{b}}_1}\xspace}
\providecommand{\sbtwo}{\ensuremath{\tilde{\mathrm{b}}_2}\xspace}
\providecommand{\sLep}{\ensuremath{\tilde{\mathrm{l}}}\xspace}
\providecommand{\sLepC}{\ensuremath{\tilde{\mathrm{l}}^\mathrm{C}}\xspace}
\providecommand{\sEl}{\ensuremath{\tilde{\mathrm{e}}}\xspace}
\providecommand{\sElC}{\ensuremath{\tilde{\mathrm{e}}^\mathrm{C}}\xspace}
\providecommand{\seL}{\ensuremath{\tilde{\mathrm{e}}_\mathrm{L}}\xspace}
\providecommand{\seR}{\ensuremath{\tilde{\mathrm{e}}_\mathrm{R}}\xspace}
\providecommand{\snL}{\ensuremath{\tilde{\nu}_L}\xspace}
\providecommand{\sMu}{\ensuremath{\tilde{\mu}}\xspace}
\providecommand{\sNu}{\ensuremath{\tilde{\nu}}\xspace}
\providecommand{\sTau}{\ensuremath{\tilde{\tau}}\xspace}
\providecommand{\Glu}{\ensuremath{\mathrm{g}}\xspace}
\providecommand{\sGlu}{\ensuremath{\tilde{\mathrm{g}}}\xspace}
\providecommand{\Wpm}{\ensuremath{\mathrm{W}^{\pm}}\xspace}
\providecommand{\sWpm}{\ensuremath{\tilde{\mathrm{W}}^{\pm}}\xspace}
\providecommand{\Wz}{\ensuremath{\mathrm{W}^{0}}\xspace}
\providecommand{\sWz}{\ensuremath{\tilde{\mathrm{W}}^{0}}\xspace}
\providecommand{\sWino}{\ensuremath{\tilde{\mathrm{W}}}\xspace}
\providecommand{\Bz}{\ensuremath{\mathrm{B}^{0}}\xspace}
\providecommand{\sBz}{\ensuremath{\tilde{\mathrm{B}}^{0}}\xspace}
\providecommand{\sBino}{\ensuremath{\tilde{\mathrm{B}}}\xspace}
\providecommand{\Zz}{\ensuremath{\mathrm{Z}^{0}}\xspace}
\providecommand{\sZino}{\ensuremath{\tilde{\mathrm{Z}}^{0}}\xspace}
\providecommand{\sGam}{\ensuremath{\tilde{\gamma}}\xspace}
\providecommand{\chiz}{\ensuremath{\tilde{\chi}^{0}}\xspace}
\providecommand{\chip}{\ensuremath{\tilde{\chi}^{+}}\xspace}
\providecommand{\chim}{\ensuremath{\tilde{\chi}^{-}}\xspace}
\providecommand{\chipm}{\ensuremath{\tilde{\chi}^{\pm}}\xspace}
\providecommand{\Hone}{\ensuremath{\mathrm{H}_\mathrm{d}}\xspace}
\providecommand{\sHone}{\ensuremath{\tilde{\mathrm{H}}_\mathrm{d}}\xspace}
\providecommand{\Htwo}{\ensuremath{\mathrm{H}_\mathrm{u}}\xspace}
\providecommand{\sHtwo}{\ensuremath{\tilde{\mathrm{H}}_\mathrm{u}}\xspace}
\providecommand{\sHig}{\ensuremath{\tilde{\mathrm{H}}}\xspace}
\providecommand{\sHa}{\ensuremath{\tilde{\mathrm{H}}_\mathrm{a}}\xspace}
\providecommand{\sHb}{\ensuremath{\tilde{\mathrm{H}}_\mathrm{b}}\xspace}
\providecommand{\sHpm}{\ensuremath{\tilde{\mathrm{H}}^{\pm}}\xspace}
\providecommand{\hz}{\ensuremath{\mathrm{h}^{0}}\xspace}
\providecommand{\Hz}{\ensuremath{\mathrm{H}^{0}}\xspace}
\providecommand{\Az}{\ensuremath{\mathrm{A}^{0}}\xspace}
\providecommand{\Hpm}{\ensuremath{\mathrm{H}^{\pm}}\xspace}
\providecommand{\sGra}{\ensuremath{\tilde{\mathrm{G}}}\xspace}
\providecommand{\mtil}{\ensuremath{\tilde{m}}\xspace}
\providecommand{\rpv}{\ensuremath{\rlap{\kern.2em/}R}\xspace}
\providecommand{\LLE}{\ensuremath{LL\bar{E}}\xspace}
\providecommand{\LQD}{\ensuremath{LQ\bar{D}}\xspace}
\providecommand{\UDD}{\ensuremath{\overline{UDD}}\xspace}
\providecommand{\Lam}{\ensuremath{\lambda}\xspace}
\providecommand{\Lamp}{\ensuremath{\lambda'}\xspace}
\providecommand{\Lampp}{\ensuremath{\lambda''}\xspace}
\providecommand{\spinbd}[2]{\ensuremath{\bar{#1}_{\dot{#2}}}\xspace}

\providecommand{\MD}{\ensuremath{{M_\mathrm{D}}}\xspace}% ED mass
\providecommand{\Mpl}{\ensuremath{{M_\mathrm{Pl}}}\xspace}% Planck mass
\providecommand{\Rinv} {\ensuremath{{R}^{-1}}\xspace} 

%% file: Introduction.tex
%%%%%%%%%%%%%%%%%
\section{Introduction\label{sec:intro}}
%%%%%%%%%%%%%%%%%

Heavy-flavour and quarkonium production 
at hadron colliders 
provides an important test of the theory of Quantum Chromodynamics (QCD). 
The production of \JPsi mesons 
%%%at hadron colliders 
occurs in three ways: prompt \JPsi produced directly in the proton-proton collision, prompt \JPsi produced indirectly (via decay of heavier charmonium states such as $\chi_c$), and non-prompt \JPsi from the decay of a b hadron. 
This paper presents the first measurement of the differential inclusive, prompt and non-prompt (b hadron) 
\JPsi production cross sections
 in pp collisions at a centre-of-mass energy of 7~TeV, in the rapidity range $|y|<2.4$, by the Compact Muon Solenoid (CMS) experiment.  

Despite considerable progress in recent years~\cite{yellow,lansberg,kramer}, 
quarkonium production remains puzzling and none of the existing theoretical models 
satisfactorily describes the prompt \JPsi differential cross 
section~\cite{kramer,Abe:1997jz,bib-cdfjpsi} and 
polarization~\cite{bib-cdfpol_jpsi} measured at the Tevatron~\cite{bib-faccioli}. 
Measurements at the Large Hadron Collider (LHC) will contribute to the 
clarification of the quarkonium production mechanisms by providing
differential cross sections in wider rapidity ranges and up to higher 
transverse momenta than was  previously possible, and with corresponding 
measurements of quarkonium polarization.
Cross-section results are largely dependent on the \JPsi polarization, as 
different polarizations cause different muon momentum spectra  
in the laboratory frame. Given the sizeable
extent of this effect, for prompt \JPsi mesons (where the
polarization is presently not well described by the theoretical models) 
we choose 
to quote final results for different polarization scenarios, instead of 
treating this effect as a source of systematic uncertainty. 

Non-prompt \JPsi production can be directly related to b-hadron production, 
leading to a measurement of the b-hadron cross section in pp collisions. 
Past discrepancies between the Tevatron results (both from 
inclusive~\cite{bib-cdfjpsi}
and exclusive~\cite{cdfxs_bplus} measurements) and the next-to-leading-order (NLO) QCD theoretical calculations,
were recently resolved using the fixed-order next-to-leading-log (FONLL)
approach and updated measurements of the $\mathrm b \rightarrow \JPsi$ 
fragmentation and decay~\cite{cacciar1, cacciar2}. 
Measured cross-section values and spectra are also found to be in agreement 
with Monte Carlo generators following this approach,
such as MC@NLO~\cite{frix1, frix2}.  

The paper is organized as follows. Section 2 describes the CMS detector. 
Section 3 presents the data collection, the event trigger and selection, the \JPsi~reconstruction, and the Monte Carlo simulation. 
Section 4 is devoted to the evaluation of the detector acceptance and efficiencies to detect \JPsi~events in CMS. In Section 5 the 
measurement of the \JPsi inclusive cross section is reported. In Section 6 
the fraction of \JPsi events from b-hadron decays is 
derived, and cross-section results are presented both for prompt \JPsi 
production and for \JPsi production from b-hadron decays. 
Section 7 presents comparisons between the measurements and  model calculations.

%% file: CMS.tex
%%%%%%%%%%%%%%%%%%%%%%%%%%
\section{The CMS detector}
%%%%%%%%%%%%%%%%%%%%%%%%%% 

The central feature of the CMS apparatus is a superconducting solenoid, of 6~m internal diameter, providing a field of 3.8~T. 
Within the field volume are the silicon pixel and strip tracker, the crystal electromagnetic calorimeter and the brass/scintillator hadron calorimeter. 
Muons are detected by three types of gas-ionization detectors embedded in the steel return yoke:
Drift Tubes (DT), Cathode Strip Chambers (CSC), and Resistive Plate Chambers (RPC).
The measurement covers the pseudorapidity window $|\eta|< 2.4$, where $\eta = - \ln [\tan (\theta / 2)]$ and 
the polar angle $\theta$ is measured from the $z$-axis, which points along the counterclockwise
beam direction.
The silicon tracker is composed of pixel detectors (three barrel
layers and two forward disks on each side of the detector, made of
66~million $100\times150$~$\mu$m$^2$ pixels) followed by microstrip detectors
(ten barrel layers plus three inner disks and nine forward disks on each side of the
detector, with 10~million strips of pitch between 80 and 184~$\mu$m).
Thanks to the strong magnetic field and the high granularity of the silicon tracker, the
transverse momentum, \pt, of the muons matched to reconstructed
tracks is measured with a resolution of about 1\,\%  for the typical muons used in this analysis.
The silicon tracker also provides the primary vertex position, with $\sim$\,20~$\mu$m accuracy. 
The first level (L1) of the CMS trigger system, composed of custom
hardware processors, uses information from the calorimeters and muon
detectors to select the most interesting events.  The High Level
Trigger (HLT) further decreases the rate before data storage.
A much more  detailed description of the CMS detector can be found elsewhere~\cite{JINST}.

%% file: DataSelection.tex
%%%%%%%%%%%%%%%%%%%%%%
\section{Data sample and event reconstruction}
\subsection{Event selection}\label{sec:data-selection}
%%%%%%%%% What data are analyzed.
The analysis is based on  a data sample recorded by the CMS detector in pp collisions at a
centre-of-mass energy of 7 TeV. The sample corresponds to a total integrated luminosity of
$314\pm  34$~nb$^{-1}$~\cite{bib-lumi}.
During this data taking period,
there were
1.6 pp collisions per bunch crossing, on average. \JPsi mesons are reconstructed in the
$\mu^+\mu^-$ decay channel.
The event selection requires good quality data from the
tracking,
muon, and luminosity detectors, in addition to good trigger conditions.

The
analysis is based on events triggered by a double-muon trigger that requires the detection of two
independent muon segments at L1, without any further processing at the HLT. All three muon systems, DT, CSC and
RPC, take part in the trigger decision.
The coincidence of two muon signals, without any cut on \pt,
is enough to keep the trigger rate reasonably low at the instantaneous
luminosities of the LHC start-up.

Events not coming from pp collisions, such as those from beam-gas interactions or beam-scraping in the transport system near the interaction point, which produce a large activity in the pixel detector, are removed
by requiring
a good primary vertex to be reconstructed~\cite{bib-vertex}.

\subsection{Monte Carlo simulation}

Simulated events are used to tune the selection criteria,
to check the agreement with data, to compute the acceptance, and to derive
corrections to the efficiencies (Section~\ref{sec:Acc_eff}).
Prompt \JPsi mesons have been simulated using Pythia 6.421~\cite{bib-PYTHIA},
which generates events based on the leading-order color-singlet and color-octet mechanisms,
with non-relativistic QCD (NRQCD) matrix elements tuned by comparing calculations with CDF data~\cite{kramer,marianne}.
Color-octet states undergo a shower evolution.
Simulated events with b-hadron pairs were also generated with Pythia and the b hadrons decayed inclusively into \JPsi using the EvtGen package~\cite{bib-evtgen}.
Final-state bremsstrahlung was
  implemented using PHOTOS~\cite{bib-photos1,bib-photos2}.

The generated events were passed through the GEANT4-based~\cite{bib-GEANT4} detector simulation and processed with
the same reconstruction program as used for collision events.
The detector simulation includes the trigger, as well as the effects of the finite precision of alignment and calibration, as determined using LHC collision data and cosmic-ray muon events \cite{bib-trackeralignment}.

\subsection{Offline muon reconstruction}\label{sec:muon_reco}

In this analysis, muon candidates are defined as tracks reconstructed in the silicon tracker which are associated with
a compatible signal in the muon chambers.

Two different muon reconstruction algorithms are considered~\cite{bib-muonreco}.
The first one provides high-quality and high-purity muon reconstruction for tracks with
\pt$\gtrsim 4$~\GeVc  in the central pseudorapidity region ($|\eta|\lesssim 1.3$) and \pt$\gtrsim 1$~\GeVc in the forward region;
these muons are referred to as {\it Global Muons}.
The second muon reconstruction algorithm achieves a better reconstruction efficiency at lower momenta; these muons are  referred to as {\it Tracker Muons}.
There is an overlap between these two reconstruction methods.
If a muon is reconstructed by both algorithms,
it is assigned to the Global Muon category alone,  making the two categories exclusive.
Global Muons have a higher reconstruction purity.
In both cases, the track momentum is determined by the fit in the silicon tracker.

To reduce muon backgrounds, mostly from decays in flight of
kaons and pions, and to ensure good quality reconstructed tracks,
muon tracks are required to pass the following requirements:
they must have at least 12 hits in the tracker,
at least two of which are required to be in the  pixel layers, a track fit with a $\chi^2$ per degree of freedom smaller than four,
 and must pass within a cylinder of radius 3~cm and length 30~cm centered at the primary vertex and parallel to the beam line.
 If two (or more) tracks are close to each other, it is  possible that the same muon segment or set of segments is associated with more than one track.
In this case the best track is selected
based on the matching between the extrapolated track and the segments in the muon
detectors.

The momentum measurement of charged tracks in the CMS detector
has systematic uncertainties due to imperfect knowledge
of the magnetic field, modeling of the detector material,
sub-detector misalignment, and biases in the
algorithms which fit the track trajectory; these effects can
shift and/or broaden the reconstructed peaks of dimuon resonances.
In addition to calibrations already applied to the data \cite{bib-magneticfield,bib-material,bib-trackeralignment},
residual effects can be determined by studying the dependence of the reconstructed dimuon peak shapes
on the muon kinematics.
The transverse momentum corrected for the residual scale distortion is parametrized as
\begin{equation}\label{eq:scale_bias}
p_\mathrm{T}^{\rm{corr}}  =  (1+ a_1 + a_2 \eta^2)  p_\mathrm{T}^{\rm{meas}},
\end{equation}
where $p_\mathrm{T}^{\rm{meas}}$ is the measured muon transverse momentum.
A
likelihood fit \cite{bib-trackermomentum}
was performed to the invariant mass shapes to minimize the difference between the
reconstructed \JPsi mass and the world-average value~\cite{bib-pdg}.
The resulting values of $a_1$ and $a_2$ are $(3.8\pm 1.9)\cdot 10^{-4}$ and $(3.0\pm0.7)\cdot  10^{-4}$, respectively.

\subsection{\JPsi~event selection}\label{sec:jpsi_selection}
%%%%%% Offline part: $J/\psi$ reconstruction}
To select the events with \JPsi~decays, muons with opposite charge are paired and their
invariant mass is computed. The invariant mass of the muon pair is required to be between
2.6 and 3.5~GeV/$c^2$.
The two muon trajectories are fitted with a common vertex constraint, and events are retained
if the fit $\chi^2$ probability is larger than 0.1\%.
This analysis uses combinations of two Global Muons, two Tracker Muons, and one Global and
one Tracker Muon.  On average, 1.07 \JPsi
combinations were found per selected dimuon event.
In case of multiple combinations in the same event, the one with
the purest muon content is chosen.
If there are two or more dimuon candidates
of the same type (Global-Global,
Global-Tracker, or Tracker-Tracker) the one of highest \pt is chosen.

\begin{figure}[h!]
\centering
{
\includegraphics[width=7.5cm]{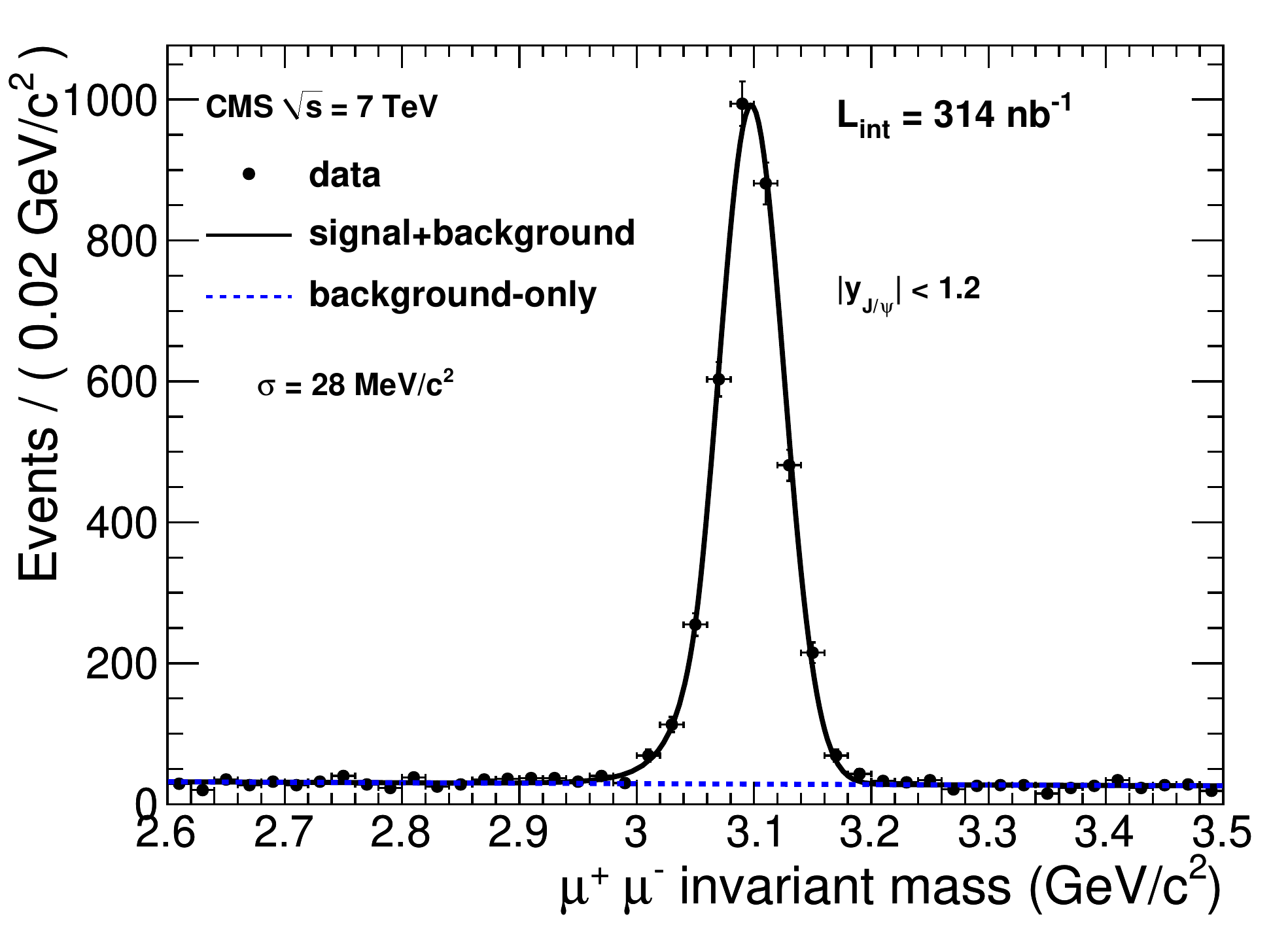}
\includegraphics[width=7.5cm]{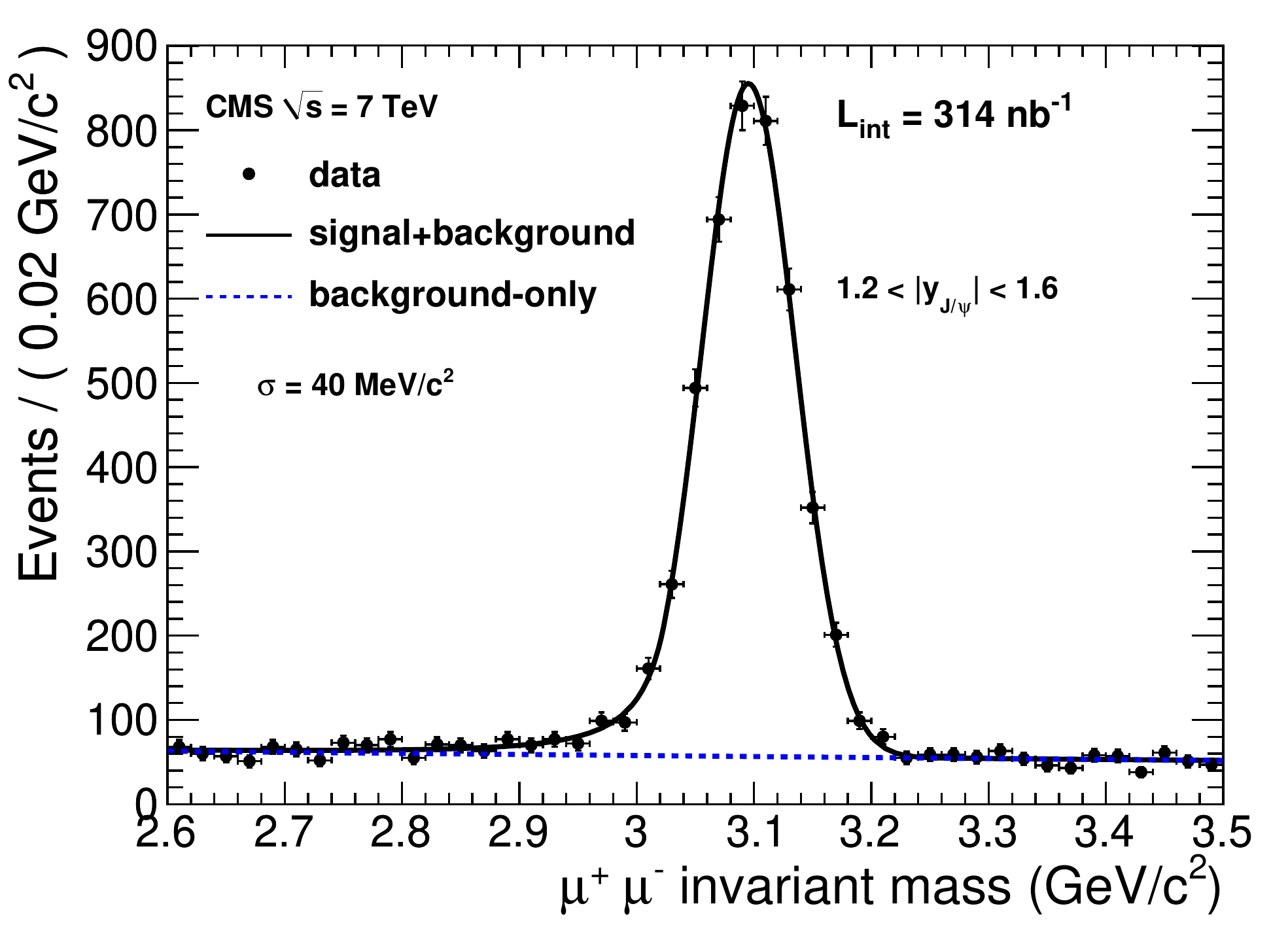}
\includegraphics[width=7.5cm]{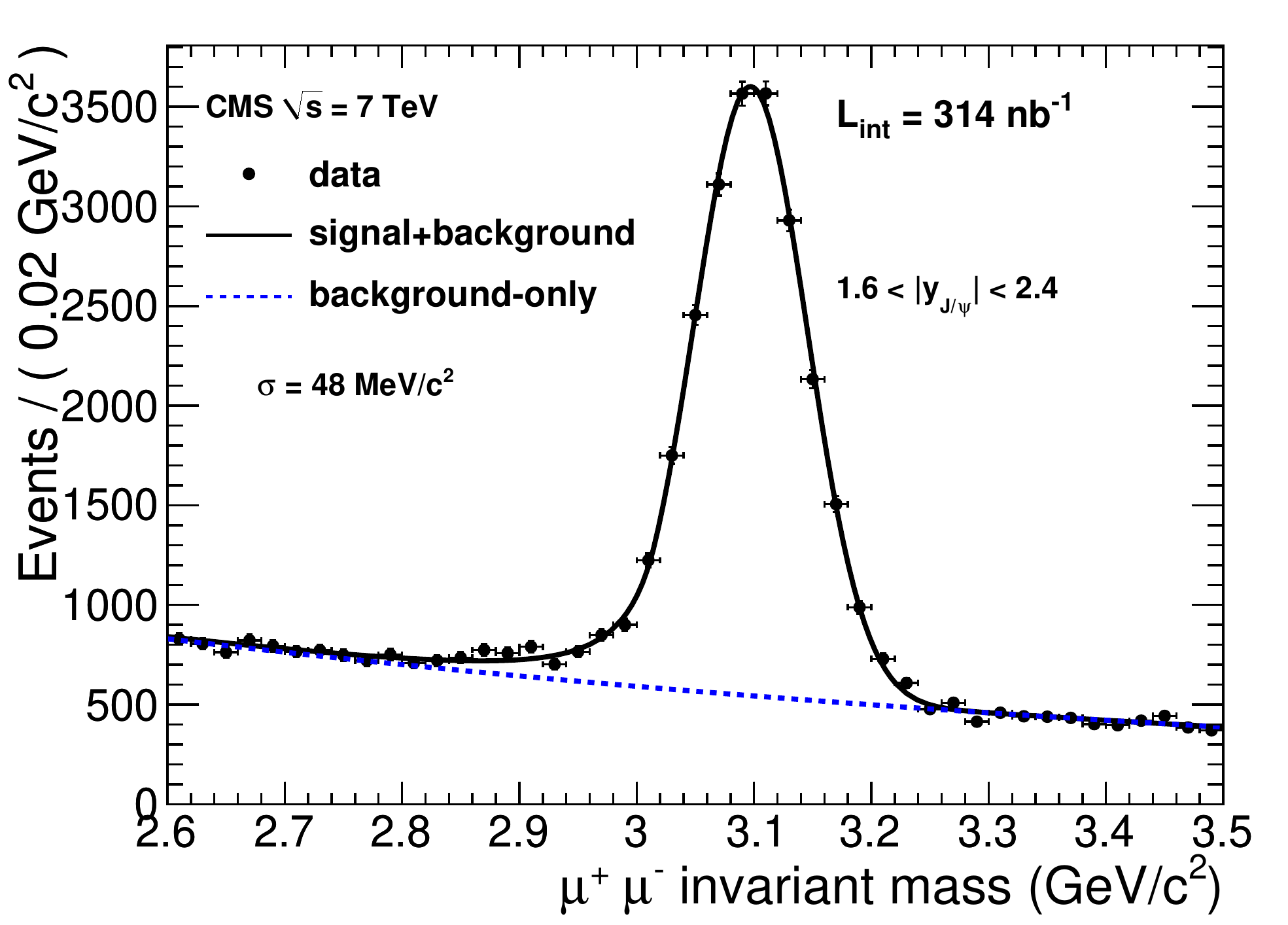}
}
\caption{
Opposite-sign  dimuon invariant mass distributions in three \JPsi rapidity ranges,
fitted with a Crystal Ball function plus an exponential (Section~\ref{inclusive}).
The poorer dimuon mass resolution at forward
rapidity is caused by the smaller lever arm of the muon tracks.
}\label{fig:massdata}
\end{figure}

The opposite-sign dimuon mass spectrum is shown in Fig.~\ref{fig:massdata} for three different \JPsi rapidity ranges.
About 27\,000 \JPsi candidates have been reconstructed, of which about 19\% are in the two-Global-Muon
category, 54\% in the Global-Tracker-Muon category, and the remaining in the two-Tracker-Muon category.

%% file: Acceptance_efficiency.tex
\section{Acceptance and Efficiency}\label{sec:Acc_eff}
%The observed number of \JPsi~ events is corrected for the detector acceptance and 
%reconstruction efficiency.
\subsection{Acceptance}

The acceptance reflects the finite geometrical coverage of the CMS detector 
and the limited kinematical reach of the muon trigger and reconstruction systems,
constrained by the thickness of the material in front of the muon detectors and by the track curvature 
in the magnetic field.

The \JPsi~acceptance $A$ is defined as the fraction of detectable 
${\mbox\JPsi}\to\mu^+\mu^-$ decays, 
as a function of the dimuon transverse momentum \pt and rapidity $y$,

\begin{equation}
%A_\alpha(\pt, y)=\frac{N_{\alpha, \mbox{\small{det}}}(\pt, y)}{N_{\alpha,\mbox{\small{gen}}}(\pt, y)}
A(\pt, y;\lambda_\theta)=\frac
{N_{\mbox{\small{det}}}(\pt, y;\lambda_\theta)}
{N_{\mbox{\small{gen}}}(\pt, y;\lambda_\theta)} \quad, 
\label{eq:acceptance}
\end{equation}
where 
$N_{\mbox{\small{det}}}$
is the number of detectable \JPsi~events in a given (\pt, $y$) bin, expressed in terms of the dimuon variables after detector smearing, and 
$N_{\mbox{\small{gen}}}$
is the corresponding total number of generated \JPsi~events in the Monte Carlo simulation. 
The parameter $\lambda_\theta$ reflects the fact that the acceptance is
computed for various polarization scenarios, as explained below.
The large number of simulated events available allows the use
of a much smaller bin size
for determining $A$ than
what is used for the cross-section measurement.

The criteria for detecting the muons coming from the \jpsi~decay is 
that both muons should be within the geometrical acceptance of the muon detectors and have
enough momentum to reach the muon stations. 
The following kinematic cuts, defining the acceptance region, are chosen 
so as to guarantee a single-muon detection probability exceeding about 
10\%:
\begin{eqnarray*}
\ptmu>3.3~ \GeVc  & \mbox{for} & |\etamu|<1.3 \quad;  \\
p^\mu>2.9~ \GeVc  & \mbox{for} & 1.3<|\etamu|<2.2 \quad ; \\
\ptmu>2.4~ \GeVc  & \mbox{for} & 2.2<|\etamu|<2.4 \quad .
\end{eqnarray*}
To compute the acceptance, \JPsi~events are generated with no cut on \pt~and  within a rapidity region extending beyond the muon detector's coverage. 

\begin{figure}[h!]
\centering

{\includegraphics[width=7cm]{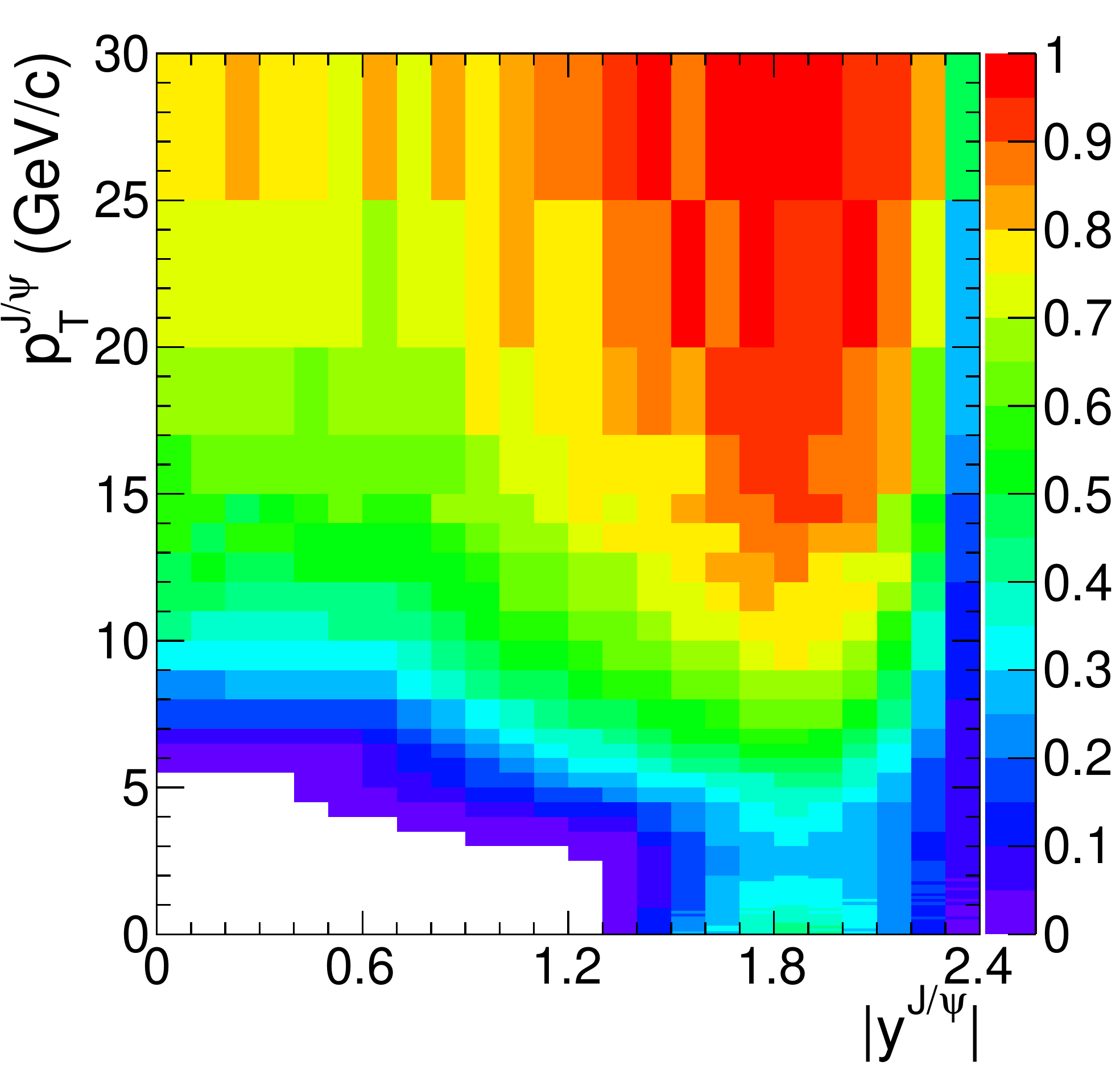}
\includegraphics[angle=90,width=7cm]{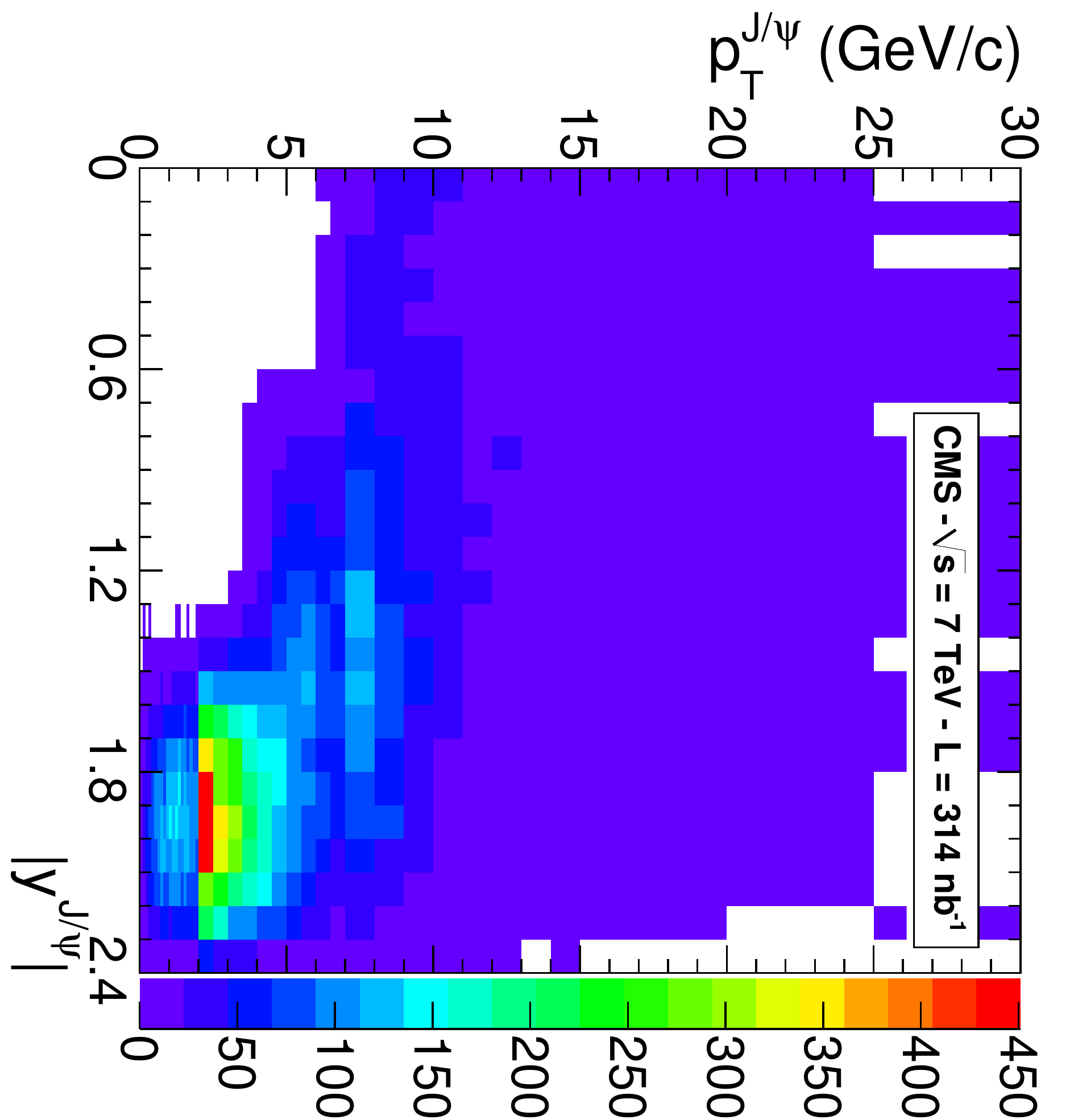}}
\caption{Left: Acceptance as a function of the \JPsi \pt~and rapidity. 
Right: Number of muon pairs within $\pm$\,100~MeV/$c^2$ of the nominal \JPsi~mass, in bins of \pt~and $|y|$.}
\label{fig:acceptance}
\end{figure}

The acceptance as a function of \pt~and $|y|$ is shown in the left plot of Fig.~\ref{fig:acceptance} 
for the combined prompt and non-prompt \JPsi~mesons, with the prompt component decaying isotropically, 
corresponding to unpolarized production. The right plot of Fig.~\ref{fig:acceptance} displays
the \pt~and $|y|$ distribution of muon pairs measured with an invariant mass
within $\pm$\,100~MeV/$c^2$ of the known \JPsi~mass~\cite{bib-pdg}. 

Systematic uncertainties on the acceptance have been investigated, as described in the following paragraphs.

\begin{itemize}
\item {\bf Final-state radiation}. At the generator level, the dimuon momentum may differ from the \JPsi momentum, due to final-state radiation (FSR). The difference between the acceptance computed using the dimuon system or the \JPsi~variables in Eq.~\ref{eq:acceptance} is taken as a systematic uncertainty.   
\item {\bf Kinematical distributions}. Different spectra of the generated \JPsi might produce different acceptances. The difference between using the Pythia spectra and other
theoretical calculations (mentioned in Section~\ref{sec:comparison}) is taken as a systematic uncertainty. 
\item {{\bf b-hadron fraction and polarization}}. 
The \JPsi~mesons produced in b-hadron decays can, in principle, have a
different acceptance 
with respect to the prompt ones, due to their different momentum spectra,  leading
to an uncertainty coming from the unknown proportion of b hadrons in the inclusive sample. 
This fraction has been varied in the Monte Carlo simulation by 20\%, the average accuracy of the measurement performed here (presented in Section~\ref{sec:bfraction}); the difference between the two acceptances is taken as an estimate of this uncertainty.
For non-prompt \JPsi~mesons
the b-hadron events are
generated with the \JPsi~polarization as
measured by the BaBar experiment~\cite{babarpol}, and
the corresponding systematic uncertainty is evaluated by taking the difference with respect to the one predicted by EvtGen. 
\item {\bf \pt~calibration and resolution}. A difference between the muon momentum scale in
data and simulated events would lead to a different acceptance.  
The muon transverse momenta have been calibrated as described 
in Section~\ref{sec:muon_reco}. The maximum residual bias remaining after 
the calibration is estimated to be 0.05\%.
As a conservative estimate, a bias equivalent to 
%plus or minus the 
this residual uncertainty 
%%%on the data 
is applied to the simulated muon momenta. The change in the recomputed acceptance is 
%%%recomputed with those shifted momenta, and the difference 
taken as a systematic uncertainty.
Similarly, a difference in the momentum resolution between data and simulated events would also give a different
%%%systematic uncertainty on the 
acceptance.
The acceptance has been computed with simulated muon momenta smeared according to the resolution
measured in data~\cite{bib-trackermomentum}
%%%, adding a contribution equal to plus or minus its uncertainty. The 
and the difference is taken as a systematic uncertainty.
\end{itemize}

Finally, the distribution of the $z$ position of the pp interaction point 
could in principle influence the acceptance. Several Monte Carlo samples 
of \JPsi 
mesons have been generated, each coming from different positions along the 
beam line (between $-10$ and $+10$ cm with respect to the centre of the collision region) and a negligible variation of the acceptance has been found.

\subsection{Efficiency}

The single-muon efficiency is computed using the \textit{Tag-and-Probe} method~\cite{bib-muonreco,bib-trackingefficiency}.
The combined  trigger and offline-reconstruction efficiency for a single muon
is measured with a data sample collected with looser trigger requirements and is defined as
\begin{equation}
\effic(\mu)  = \effic_{\mbox{\small track}} \cdot \effic_{\mbox{\small id $|$ track}} \cdot \effic_{\mbox{\small trig $|$ track+id}} \quad ,
\end{equation}
where $ \effic_{\mbox{\small track}} $ is the tracking efficiency, $ \effic_{\mbox{\small id $|$ track}} $ is the muon identification efficiency in the muon systems for a tracker-reconstructed muon,
and finally $ \effic_{\mbox{\small trig $|$ track+id}}$ is the probability for an offline reconstructed muon to have also fired the trigger.

The tracking efficiency is constant
in the momentum range defined by the acceptance cuts,
and it varies only slightly in the $\phi-\eta$ plane~\cite{bib-trackingefficiency}.
The muon identification and trigger efficiencies have a stronger
\ptmu~and $|\etamu|$ dependence, 
which is
mapped with a finer granularity
(nine to twelve
\ptmu~and five $|\etamu|$ bins). 

The efficiency to detect a given \JPsi~event is thus dependent on the 
value of the muon-pair kinematic variables, and is given by 
\begin{equation}
\epsilon (\JPsi) = \epsilon ({\mu^+})  \cdot \epsilon ({\mu^-}) \cdot(1+\rho)
\cdot \epsilon_{\mbox{\small vertex}}\quad.
\label{eq:jpsi_eff}
\end{equation}

The factor $\rho$ represents a correction to the factorization hypothesis
and is evaluated from the Monte Carlo simulation. The 
non-vanishing values of $\rho$, varying between $-0.19$ and $0.30$,
are mainly due to the relatively large bin sizes used to determine the muon efficiencies. 

The efficiency for the two muon tracks to be consistent 
with coming from a common vertex (Section~\ref{sec:jpsi_selection}), 
$\epsilon_{\mbox{\small vertex}}$,
is measured to be $(98.35\pm 0.16)\%$, by comparing the number of two-Global-Muon
 combinations within $\pm$\,100~MeV/$c^2$
 of the nominal \JPsi~mass with and without the common vertex requirement. Given the precision of this estimate,
the corresponding systematic uncertainty can be neglected.
The following systematic uncertainties on the \JPsi~ efficiency are considered:
\begin{itemize}
\item {\bf $\rho$ factor}. Any variation of the muon spectrum within each large bin may lead to a different value of $\rho$.
By reweighting the Pythia Monte Carlo simulation, we vary the \JPsi~\pt~spectrum to reproduce different theoretical predictions (Section~\ref{sec:comparison}), and
take the largest variation as the systematic uncertainty on $\rho$.

\item {\bf Muon efficiency}. The statistical uncertainty on each muon efficiency 
is propagated using toy Monte Carlo experiments, 
and the r.m.s.\ of the newly computed \JPsi efficiencies are assigned as systematic uncertainties. The largest systematic errors are in the bins with less events or in those
where the background is largest.
When selecting the tag muon, the Tag-and-Probe method produces a slight bias 
on the kinematics of the probe muon, hence a small difference arises between 
the measured single-muon efficiencies and those of an unbiased sample. 
This small effect is studied in the Monte Carlo simulation and corrected for.
The whole correction is conservatively taken as a systematic uncertainty 
on the efficiencies and summed in quadrature with the 
statistical uncertainty.
%%%, to obtain the total uncertainty on the muon efficiencies. 
\end{itemize}

%% file: Inclusive.tex
%%%%%%%%%%%%%%%%%%
\section{Inclusive \texorpdfstring{\JPsi}{J/Psi}~cross section}\label{inclusive}
%%%%%%%%%%%%%%%%%%
The measurement of the inclusive \pt~differential cross section is based on the equation
\begin{equation}
\frac{d^2\sigma}{dp_\mathrm{T}dy}(\JPsi) \cdot \mbox{BR}(\JPsi\rightarrow \mu^{+}\mu^{-})=\frac{N_{\mbox{corr}}(\JPsi)}{\int{}
Ldt\cdot\Delta p_\mathrm{T}\cdot\Delta y}\quad ,
\end{equation}
where $N_{\mbox{corr}}(\JPsi)$ is the \JPsi~yield, corrected for the \JPsi~acceptance and selection
efficiency, in a given transverse momentum-rapidity  bin,
$\int{}Ldt$ is the integrated luminosity, $\Delta p_\mathrm{T}$ and $\Delta y$ are the sizes
of the \pt~and rapidity bins, and $\mbox{BR}(\JPsi\rightarrow \mu^{+}\mu^{-})$ is
the branching ratio of the \JPsi~decay into two muons.

\subsection{\JPsi~yields}\label{sec:yields}
The corrected yield, $N_{\mbox{corr}}(\JPsi)$, is determined in two steps. First, in each rapidity and \pt~bin  an unbinned maximum likelihood fit to the $\mu^+\mu^-$ invariant mass distribution is performed.
The resulting yield is then corrected by a factor that takes into account
the average acceptance ($A$) and detection efficiency ($\epsilon$) in the
bin under consideration.

In the mass fits, the shape assumed for the signal is a Crystal Ball function \cite{bib-crystalball}, which takes into account the detector resolution as well as the radiative tail from bremsstrahlung. The shape of the underlying continuum is described by an exponential. Table~\ref{maintable_d} lists the \JPsi~uncorrected signal yields and the corresponding statistical
uncertainties from the fit, for the chosen bins.
\begin{table}[h!]
\caption{\small{Uncorrected event yield (with its statistical error from the fit) in each \pt bin, together with the average
    acceptance times efficiency (computed in the unpolarized production scenario).
}\label{maintable_d}}
\begin{center}
{\small
\begin{tabular}{lrc||lrc}
\hline
\ptjpsi~ (\GeVc) & Yield & $\left\langle 1/(A\epsilon)\right\rangle^{-1}$ &\ptjpsi~ (\GeVc) & Yield & $\left\langle 1/(A\epsilon)\right\rangle^{-1}$ \\
\hline
& &  & \multicolumn{3}{c}{$1.6<|y|<2.4$}\\\cline{4-6}
& & &                                                    $0.00-0.50$   & $695.6 \pm 40.7$ & $0.075 \pm 0.008$ \\\cline{1-3}
\multicolumn{3}{c||}{$|y|<1.2$} &                        $0.50-0.75$   & $829.3 \pm 44.7$ & $0.079 \pm 0.010$ \\\cline{1-3}
$6.5-8.0  $ & $726.5 \pm 28.3$ & $0.084 \pm 0.005$ &     $0.75-1.00$   & $1006.0 \pm 48.8$ & $0.078 \pm 0.010$\\
$8.0-10.0 $ &$868.1 \pm 30.7$  & $0.178 \pm 0.005$ &     $1.00-1.25$   & $1216.8 \pm 52.8$ & $0.079 \pm 0.010$\\
$10.0-12.0$ &$513.2 \pm 23.5$  & $0.288 \pm 0.008$ &     $1.25-1.50$   & $1232.9 \pm 53.7$ & $0.077 \pm 0.008$\\
$12.0-30.0 $ &$636.0 \pm 26.1$ & $0.405 \pm 0.008$ &    $1.50-1.75$   & $1252.9 \pm 50.3$ & $0.075 \pm 0.008$\\\cline{1-3}
 &  &  &   $1.75-2.00$   & $1132.7 \pm 57.5$ & $0.074 \pm 0.006$\\
 &  &  &   $2.00-2.25$   & $1122.7 \pm 55.0$ & $0.071 \pm 0.006$ \\\cline{1-3}
\multicolumn{3}{c||}{$1.2<|y|<1.6$} &                   $2.25-2.50$   & $899.9 \pm 39.4$ & $0.074 \pm 0.006$ \\ \cline{1-3}
$2.0-3.5$ & $414.9 \pm 38.0$ & $0.016 \pm 0.001$   &    $2.50-2.75$   & $903.3 \pm 72.4$ & $0.075 \pm 0.004$ \\
$3.5-4.5$ & $401.7 \pm 23.2$ & $0.035 \pm 0.004$   &    $2.75-3.00$   & $757.6 \pm 36.2$ & $0.077 \pm 0.005$ \\
$4.5-5.5$ & $618.9 \pm 28.9$ & $0.086 \pm 0.004$   &    $3.00-3.25$   & $756.1 \pm 35.7$ & $0.082 \pm 0.005$ \\
$5.5-6.5$ & $690.9 \pm 34.0$ & $0.167 \pm 0.005$   &    $3.25-3.50$   & $703.6 \pm 33.6$ & $0.084 \pm 0.004$ \\
$6.5-8.0$ & $712.0 \pm 28.0$ & $0.247 \pm 0.006$   &    $3.50-4.00$   & $1150.2 \pm 40.0$ & $0.092 \pm 0.005$ \\
$8.0-10.0$ & $463.7 \pm 23.3$ & $0.334 \pm 0.009$  &    $4.00-4.50$   & $991.8 \pm 35.8$ & $0.100 \pm 0.004$ \\
$10.0-30.0$ & $406.2 \pm 22.4$ & $0.445 \pm 0.010$ &    $4.50-5.50$   & $1441.4 \pm 42.6$ & $0.117 \pm 0.005$ \\ \cline{1-3}
& & &                                                   $5.50-6.50$   & $993.0 \pm 34.7$ & $0.157 \pm 0.008$ \\
& & &                                                   $6.50-8.00$   & $900.6 \pm 35.1$ & $0.193 \pm 0.008$ \\
& & &                                                   $8.00-10.00$  & $604.3 \pm 26.8$ & $0.250 \pm 0.007$ \\
& & &                                                   $10.00-30.00$ & $462.6 \pm 23.6$ & $0.309 \pm 0.010$ \\\hline

\hline
\end{tabular}
}
\end{center}
\end{table}

Different functions were used to assess systematic effects coming from the fit function chosen to model the signal and the continuum shapes.
For the signal, the Crystal Ball function was varied to a sum of a Crystal Ball and a Gaussian, while for the background a second-order polynomial was used.
The maximum difference in the result was taken as a systematic uncertainty. The uncertainty is particularly large for the low-\pt bins, where the signal purity is the smallest.

Additionally, a bias on the muon momentum scale can shift the events from one \JPsi
\pt bin to the adjacent ones. To estimate this systematic effect, a bias has been applied to the muon momenta equal to
the residual uncertainty on the scale after the calibration, as explained in Section~\ref{sec:jpsi_selection}, and a negligible variation was found.

\begin{table}[h!]
\centering
\caption{\small{Relative systematic uncertainties on the corrected yield for different \JPsi rapidity bins.
 The variation range over the different \pt bins is given. In general, uncertainties depend
only weakly on the \pt values, except for the fit function systematic uncertainty, which decreases with increasing \pt due to
the better purity of the signal. The large excursion of the muon efficiency systematic uncertainty reflects
changes in the event yield and in the signal purity among the \pt bins.}\label{syst}}
{\small
\begin{tabular}{llccc}
\hline
Affected quantity &Source &  \multicolumn{3}{c}{Relative error (\%)}\\ \cline{3-5}
 & & $|y|<1.2$ & $1.2<|y|<1.6$ & $1.6<|y|<2.4$\\
\hline
Acceptance & FSR & $0.8-2.5$ & $0.3-1.6$ & $0.0-0.9$\\
& \pt calibration and resolution& $1.0-2.5$ & $0.8-1.2$ & $0.1-1.0$\\
& Kinematical distributions & $0.3-0.8$ & $ 0.6-2.6$ & $0.9-3.1$\\
& b-hadron fraction and polarization & $1.9-3.1$&$0.5-1.2$ & $0.2-3.0$ \\
\hline
Efficiency & Muon efficiency &$1.9-5.1$&$2.3-12.2$ & $2.7-9.2$\\
& $\rho$ factor &$0.5-0.9$& $0.6-8.1$&$0.2-7.1$ \\\hline
Yields & Fit function & $0.6-1.1$& $0.4-5.3$& $0.3-8.8$\\
\hline
\end{tabular}
}
\end{table}

\subsection{Inclusive \JPsi~cross section results}

The previously discussed systematic uncertainties affecting the inclusive \JPsi
cross section are listed in Table~\ref{syst}. In addition, the relative error on the luminosity determination is 11\%, and is common to all bins.
Table~\ref{tab:results_d}
reports the values of the resulting \JPsi differential
cross section, for different polarization scenarios:
unpolarized, full longitudinal polarization and full transverse
polarization
in the Collins-Soper or the helicity frames~\cite{bib-faccioli}.

\begin{figure}[h!]
\centering
\includegraphics[angle=90,width=12cm]{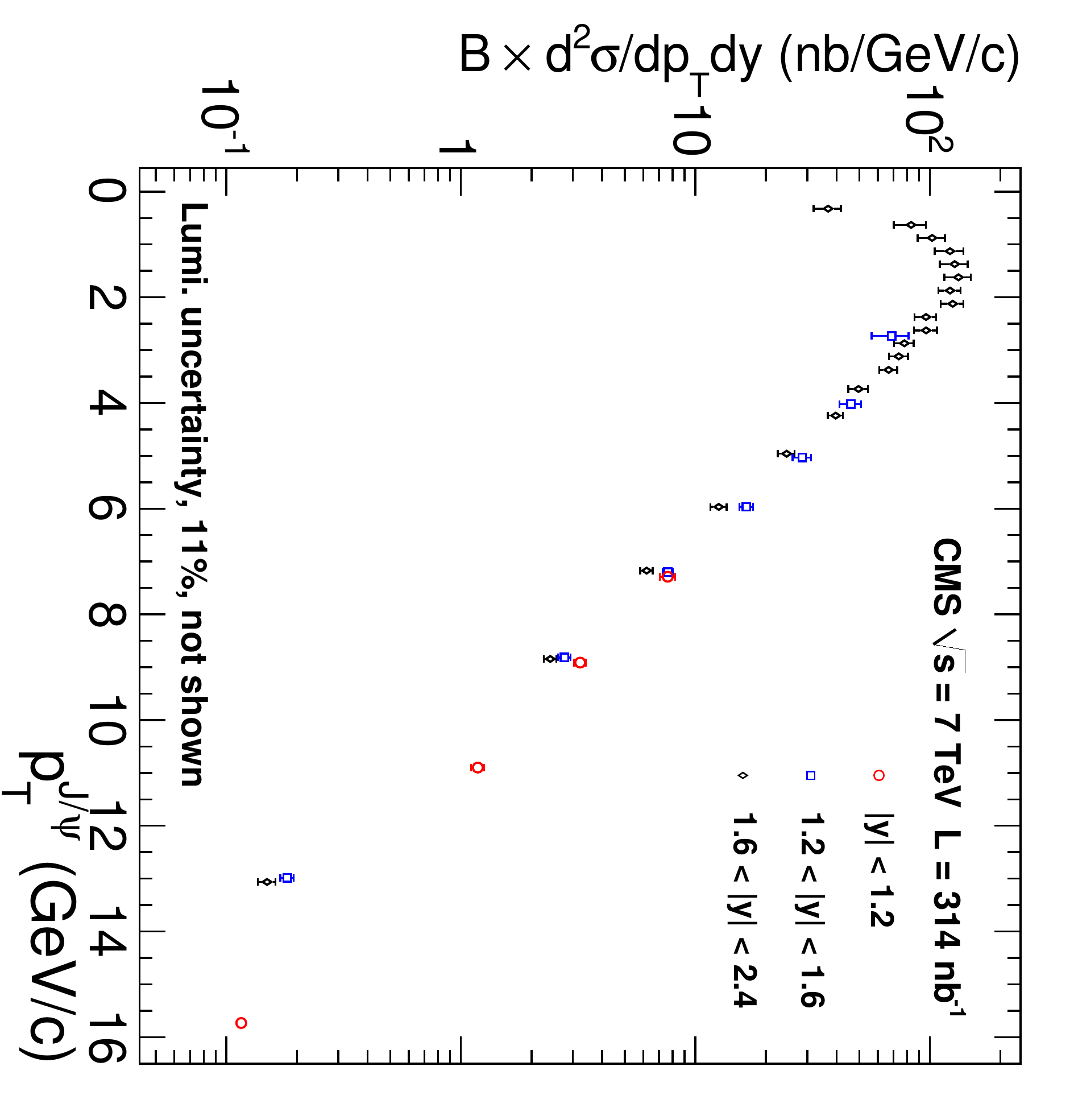}
\caption{Differential inclusive \JPsi cross section as a function of \pt for the three
  different rapidity intervals and in the unpolarized production scenario.  The errors on the ordinate values are the statistical and systematic errors added in quadrature. The 11\% uncertainty due to the luminosity determination is not shown and is common to all bins.}\label{fig:xsec}
\end{figure}

Figure~\ref{fig:xsec} shows the inclusive differential cross section
$\frac{d^2\sigma}{d\pt dy} \cdot BR(\JPsi\to\mu^+\mu^-)$
in the three rapidity ranges, showing statistical and systematic uncertainties, except the luminosity uncertainty,
added in quadrature. It should be noted that the first bin in the forward rapidity region extends down to zero \JPsi \pt.

\begin{sidewaystable}[htdp]
%\begin{table}[htdp]
\caption{Differential inclusive cross sections and  average \pt in the bin, for each prompt \JPsi polarization scenario considered: unpolarized ($\lambda_\theta=0$), full longitudinal polarization
($\lambda_\theta =-1$) and full transverse polarization ($\lambda_\theta=+1$)
in the Collins-Soper (CS) or the helicity (HX) frames~\cite{bib-faccioli}.
For the unpolarized case, the first error is statistical and the second is systematic; for the others the total error is given.}\label{tab:results_d}
\begin{center}
{\small
\begin{tabular}{lcccccc}\hline
 \ptjpsi~ & $\langle\ptjpsi\rangle$& \multicolumn{5}{c}{$\frac{d^2\sigma}{d\pt dy}\cdot\mbox{BR}(\JPsi\to\mu^+\mu^-)$  (nb/\GeVc)} \\
  (\GeVc)  &  (\GeVc) & $\lambda_\theta=0$   & $\lambda^{CS}_\theta=-1$ & $\lambda^{CS}_\theta=+1$ & $\lambda^{HX}_\theta=-1$ & $\lambda^{HX}_\theta=+1$ \\
\hline
\multicolumn{7}{c}{$|y| < 1.2$}\\
\hline
$6.50-8.00$ & $7.29$ & $7.63 \pm 0.30 \pm 0.97$ & $9.28 \pm 1.20$ & $6.99 \pm 0.91$ & $5.70 \pm 0.74$ & $9.14 \pm 1.20$ \\
$8.00-10.00$ & $8.91$ & $3.23 \pm 0.11 \pm 0.38$ & $3.81 \pm 0.47$ & $3.00 \pm 0.37$ & $2.45 \pm 0.30$ & $3.85 \pm 0.48$ \\
$10.00-12.00$ & $10.90$ & $1.18 \pm 0.05 \pm 0.14$ & $1.35 \pm 0.17$ & $1.10 \pm 0.14$ & $0.93 \pm 0.12$ & $1.37 \pm 0.17$ \\
$12.00-30.00$ & $15.73$ & $0.116 \pm 0.005 \pm 0.013$ & $0.130 \pm 0.016$ & $0.110 \pm 0.013$ & $0.096 \pm 0.012$ & $0.129 \pm 0.016$ \\
\hline
\multicolumn{7}{c}{$1.2 < |y| < 1.6$}\\
\hline
$2.00-3.50$ & $2.73$ & $68.8 \pm 6.3 \pm 13.0$ & $50.4 \pm 9.9$ & $84.6 \pm 19.0$ & $50.5 \pm 9.9$ & $84.5 \pm 19.0$ \\
$3.50-4.50$ & $4.02$ & $46.1 \pm 2.7 \pm 6.5$ & $37.3 \pm 5.7$ & $52.8 \pm 8.4$ & $33.9 \pm 5.2$ & $56.4 \pm 8.8$ \\
$4.50-5.50$ & $5.03$ & $28.6 \pm 1.3 \pm 3.9$ & $28.2 \pm 4.1$ & $28.7 \pm 4.1$ & $20.8 \pm 3.0$ & $35.0 \pm 5.0$ \\
$5.50-6.50$ & $5.96$ & $16.5 \pm 0.8 \pm 2.0$ & $17.8 \pm 2.3$ & $16.0 \pm 2.0$ & $12.3 \pm 1.6$ & $20.1 \pm 2.6$ \\
$6.50-8.00$ & $7.20$ & $7.64 \pm 0.30 \pm 0.87$ & $8.71 \pm 1.10$ & $7.19 \pm 0.87$ & $5.80 \pm 0.71$ & $9.19 \pm 1.10$ \\
$8.00-10.00$ & $8.81$ & $2.76 \pm 0.14 \pm 0.32$ & $3.11 \pm 0.39$ & $2.62 \pm 0.33$ & $2.18 \pm 0.27$ & $3.24 \pm 0.41$ \\
$10.00-30.00$ & $12.99$ & $0.182 \pm 0.010 \pm 0.021$ & $0.204 \pm 0.026$ & $0.173 \pm 0.022$ & $0.151 \pm 0.019$ & $0.202 \pm 0.026$ \\
\hline
\multicolumn{7}{c}{$1.6 < |y| < 2.4$}\\
\hline
$0.00-0.50$ & $0.32$ & $36.8 \pm 2.2 \pm 6.0$ & $26.1 \pm 4.5$ & $46.5 \pm 8.0$ & $26.3 \pm 4.5$ & $45.6 \pm 7.8$ \\
$0.50-0.75$ & $0.63$ & $83.2 \pm 4.5 \pm 15.3$ & $59.5 \pm 11.3$ & $105.1 \pm 19.9$ & $60.4 \pm 11.6$ & $103.2 \pm 19.3$ \\
$0.75-1.00$ & $0.88$ & $102.3 \pm 5.0 \pm 16.9$ & $72.8 \pm 13.3$ & $128.9 \pm 23.7$ & $75.1 \pm 13.4$ & $125.0 \pm 22.8$ \\
$1.00-1.25$ & $1.13$ & $121.9 \pm 5.3 \pm 21.1$ & $87.1 \pm 14.8$ & $152.4 \pm 27.1$ & $91.11 \pm 18.2$ & $146.2 \pm 25.6$ \\
$1.25-1.50$ & $1.37$ & $127.7 \pm 5.6 \pm 21.6$ & $91.1 \pm 15.6$ & $160.1 \pm 29.3$ & $96.2 \pm 17.7$ & $152.9 \pm 28.4$ \\
$1.50-1.75$ & $1.62$ & $132.5 \pm 5.3 \pm 21.9$ & $94.7 \pm 15.8$ & $165.9 \pm 27.7$ & $101.3 \pm 16$ & $157.8 \pm 25.4$ \\
$1.75-2.00$ & $1.87$ & $121.9 \pm 6.2 \pm 17.9$ & $87.4 \pm 13.6$ & $152.1 \pm 24.7$ & $93.6 \pm 14.9$ & $143.9 \pm 23.1$ \\
$2.00-2.25$ & $2.12$ & $125.2 \pm 6.1 \pm 18.7$ & $89.8 \pm 13.9$ & $156.3 \pm 24.7$ & $97.1 \pm 14.9$ & $147.3 \pm 23.6$ \\
$2.25-2.50$ & $2.37$ & $96.3 \pm 4.2 \pm 14.1$ & $69.0 \pm 10.2$ & $120.5 \pm 18.1$ & $74.3 \pm 11$ & $114 \pm 16.8$ \\
$2.50-2.75$ & $2.63$ & $96.4 \pm 7.7 \pm 13.0$ & $69.8 \pm 11.1$ & $119.3 \pm 18.6$ & $74.8 \pm 11.8$ & $113.2 \pm 18.1$ \\
$2.75-3.00$ & $2.87$ & $77.9 \pm 3.7 \pm 10.7$ & $56.3 \pm 8.0$ & $96.4 \pm 13.9$ & $60.3 \pm 8.5$ & $91.6 \pm 13.1$ \\
$3.00-3.25$ & $3.12$ & $73.7 \pm 3.5 \pm 10.0$ & $53.8 \pm 7.7$ & $91.2 \pm 13.0$ & $57.6 \pm 8.3$ & $86.5 \pm 13.0$ \\
$3.25-3.50$ & $3.37$ & $66.7 \pm 3.2 \pm 8.8$ & $48.5 \pm 6.9$ & $82.8 \pm 12.0$ & $52.1 \pm 7.3$ & $78.3 \pm 11.0$ \\
$3.50-4.00$ & $3.74$ & $49.6 \pm 1.7 \pm 7.1$ & $37.0 \pm 5.5$ & $60.6 \pm 9.0$ & $39.0 \pm 5.8$ & $58.3 \pm 8.6$ \\
$4.00-4.50$ & $4.24$ & $39.7 \pm 1.4 \pm 5.0$ & $30.0 \pm 4.0$ & $47.3 \pm 6.3$ & $31.4 \pm 4.2$ & $46.0 \pm 6.1$ \\
$4.50-5.50$ & $4.96$ & $24.5 \pm 0.7 \pm 3.3$ & $19.3 \pm 2.6$ & $28.7 \pm 4.0$ & $19.6 \pm 2.7$ & $28.2 \pm 3.9$ \\
$5.50-6.50$ & $5.97$ & $12.6 \pm 0.4 \pm 1.7$ & $10.8 \pm 1.4$ & $14.0 \pm 1.9$ & $10.3 \pm 1.4$ & $14.3 \pm 1.9$ \\
$6.50-8.00$ & $7.17$ & $6.20 \pm 0.24 \pm 0.74$ & $5.70 \pm 0.72$ & $6.61 \pm 0.84$ & $5.13 \pm 0.65$ & $6.94 \pm 0.88$ \\
$8.00-10.00$ & $8.84$ & $2.41 \pm 0.11 \pm 0.28$ & $2.41 \pm 0.31$ & $2.44 \pm 0.31$ & $2.04 \pm 0.26$ & $2.64 \pm 0.34$ \\
$10.00-30.00$ & $13.06$ & $0.149 \pm 0.008 \pm 0.019$ & $0.155 \pm 0.021$ & $0.148 \pm 0.021$ & $0.132 \pm 0.019$ & $0.161 \pm 0.023$ \\
\hline
\end{tabular}
}
\end{center}
\end{sidewaystable}

The total cross section for inclusive \JPsi~production, obtained by integrating over \pt between 6.5 and 30\GeVc
and over rapidity between $-2.4$ and $2.4$, in the unpolarized production hypothesis, gives
 \begin{equation}
{\small \sigma(pp\rightarrow\JPsi +X)\cdot\mbox{BR}(\JPsi\to\mu^+\mu^-) = 97.5 \pm 1.5\mbox{(stat)} \pm 3.4\mbox{(syst)} \pm 10.7\mbox{(luminosity)} \mbox{ nb} .}
 \end{equation}

%% file: Bfraction.tex
%%%%%%%%%%%%%
\section{Fraction of \texorpdfstring{\JPsi}{J/Psi} from b-hadron decays}\label{sec:bfraction}
%%%%%%%%%%%%%%%%%

The measurement of the fraction of \JPsi yield coming from b-hadron decays
relies on the discrimination of the \JPsi mesons produced
away from the pp collision vertex,
determined
by the distance between the dimuon vertex and the primary vertex in the plane orthogonal to the
beam line.

The primary vertices in the event are found by performing a common fit to
tracks
for which the points of closest approach to the beam axis are
clustered in $z$, excluding the two muons forming the \JPsi candidate and using
adaptive weights to avoid biases from displaced secondary vertices.
Given the presence of pile-up,
the primary vertex in the event is not unique. According to Monte Carlo simulation studies, the
best assignment of the primary vertex is achieved by selecting the one closest
in the $z$ coordinate to the dimuon vertex.

\subsection{Separating prompt and non-prompt \JPsi}

As an estimate of the b-hadron proper decay length, the quantity
 $\ell_{\JPsi}=L_{xy}\cdot m_{\JPsi}/{\pt}$ is computed
for each \JPsi~candidate, where $m_{\JPsi}$ is
the \JPsi~mass~\cite{bib-pdg} and $L_{xy}$ is the most probable transverse decay length
in the laboratory frame~\cite{bib-mostprobable,bib-mostprobable-erratum}. $L_{xy}$ is
defined as
\begin{equation}
L_{xy} =\frac{ \mbox{\bf u}^T\sigma^{-1}\mbox{\bf x}}{\mbox{\bf u}^T\sigma^{-1}\mbox{\bf u}}\quad,
\label{ct1}
\end{equation}
where {\bf x} is the vector joining the vertex of the two muons and the
primary vertex of the event, in the transverse plane, ${\bf u}$ is the unit vector of the \JPsi \pt, and $\sigma$ is the sum of the primary and
secondary vertex covariance matrices.

To determine the fraction $f_B$ of $J/\psi$ mesons from b-hadron decays in
the data, we perform an unbinned maximum-likelihood fit in each \pt~and rapidity bin.
The dimuon mass spectrum and the $\ell_{\JPsi}$
distribution are simultaneously fit by a log-likelihood function,
\begin{equation}\label{lik}
\ln L=\sum ^{N}_{i=1}\ln F(\ell_{\JPsi}, m_{\mu\mu})\quad,
\end{equation}
where $N$ is the total number of events and $m_{\mu\mu}$ is the invariant mass
of the muon pair.
The expression for $F(\ell_{\JPsi} ,m_{\mu\mu})$ is
\begin{equation}\label{eqx}
F(\ell_{\JPsi},m_{\mu\mu})=f_{Sig}\cdot F_{Sig}(\ell_{\JPsi})\cdot
M_{Sig}(m_{\mu\mu}) + (1-f_{Sig})\cdot F_{Bkg}(\ell_{\JPsi})\cdot
M_{Bkg}(m_{\mu\mu})\quad,
\end{equation}
where:
\begin{itemize}
\item $f_{Sig}$ is the fraction of events attributed to \JPsi~sources coming from both prompt and non-prompt components;
\item $M_{Sig}(m_{\mu\mu})$ and $M_{Bkg}(m_{\mu\mu})$ are functional forms
describing the invariant dimuon mass distributions for the signal and
background, respectively, as detailed in Section~\ref{sec:yields};
\item $F_{Sig}(\ell_{\JPsi})$ and $F_{Bkg}(\ell_{\JPsi})$ are functional forms describing the $\ell_{\JPsi}$ distribution for the signal and background, respectively.

The signal part is given by a sum of prompt and non-prompt components,
\begin{equation}\label{eq:fbfp}
F_{Sig}(\ell_{\JPsi} )=f_{B}\cdot F_{B}(\ell_{\JPsi}) + (1-f_{B})\cdot
F_{p}(\ell_{\JPsi})\quad,
\end{equation}
where $f_{B}$ is the fraction of $J/\psi$  from b-hadron decays, and
 $F_{p}(\ell_{\JPsi})$ and $F_{B}(\ell_{\JPsi})$ are the $\ell_{\JPsi}$ distributions for prompt and non-prompt $J/\psi$, respectively.

As
$\ell_{\JPsi}$ should be zero in an ideal detector for prompt events,
$F_{p}(\ell_{\JPsi})$ is described simply by a resolution function.
The core of the resolution function is taken to be a double-Gaussian and its parameters are allowed to float in the nominal
fit. Since $\ell_{\JPsi}$ depends on the position of the primary vertex, an additional Gaussian component is added, to take into account possible
wrong assignments of the primary vertex; its parameters are fixed from
the Monte Carlo simulation.

The $\ell_{\JPsi}$ shape of the non-prompt component in Eq.~\ref{eq:fbfp} is given by convolving the same resolution function
with the true $\ell_{\JPsi}$ distribution of the \JPsi from
long-lived b hadrons, as given by the Monte Carlo simulation.

For the background $\ell_{\JPsi}$ distribution $F_{Bkg}(\ell_{\JPsi})$, the functional form
employed by CDF~\cite{bib-cdfjpsi} is used:
\begin{eqnarray}
\nonumber
F_{Bkg}(x) & = & (1-f_+-f_--f_{\mathrm{sym}}) R(x) + \\
\nonumber
& & [\frac{f_+}{\lambda_+}e^{-\frac{x'}{\lambda_+}}\theta(x') +
\frac{f_-}{\lambda_-}e^{\frac{x'}{\lambda_-}}\theta(-x') + \\
& & + \frac{f_{\mathrm{sym}}}{2\lambda_{\mathrm{sym}}}e^{-\frac{|x'|}{\lambda_{\mathrm{sym}}}}] \otimes R(x'-x)~,
\end{eqnarray}
where $R(x)$ is the resolution model mentioned above,
$f_i$ ($i = \{+,-,\mathrm{sym}\}$) are the
fractions of the three long-lived components with mean
decay lengths $\lambda_i$,
and $\theta(x)$ is the step function.
The effective parameters $\lambda_i$ are previously
determined with a fit to the $\ell_{\JPsi}$ distribution in the sidebands of the dimuon invariant mass
distribution, defined as the regions 2.6--2.9 and 3.3--3.5~GeV/$c^2$.
\end{itemize}

\begin{figure}[h!]
\centering
\includegraphics[width=7.5cm]{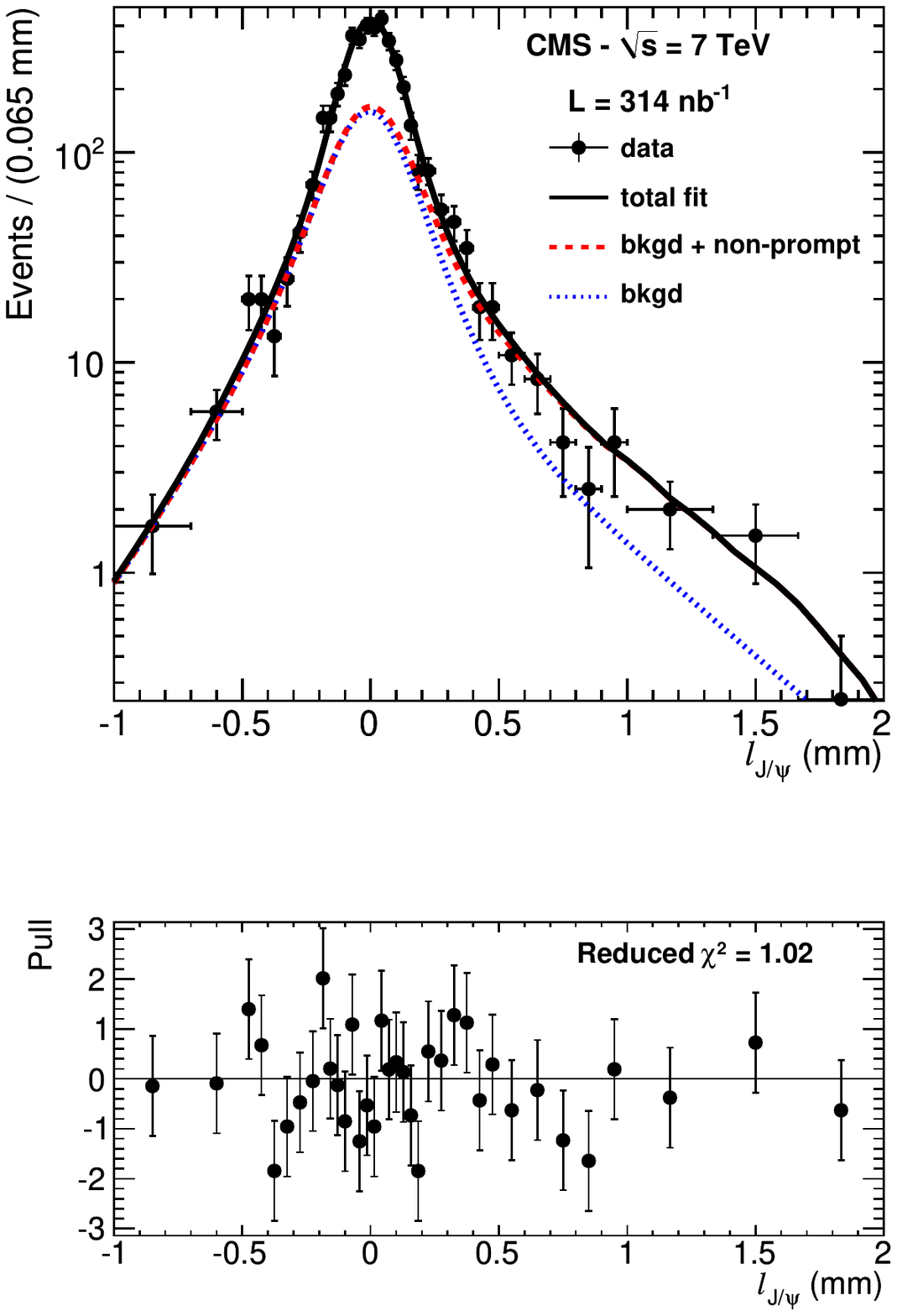}
\includegraphics[width=7.5cm]{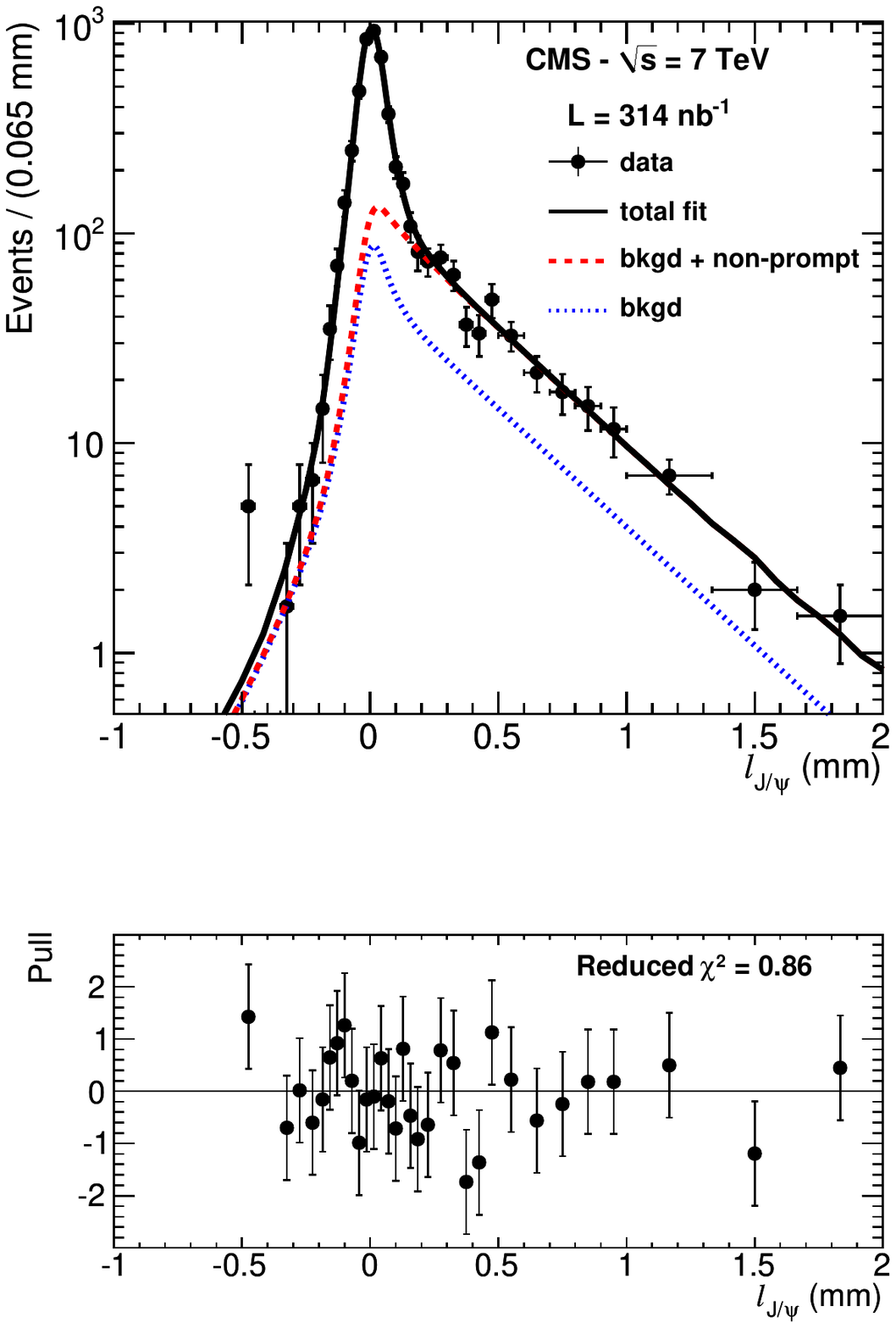}
\caption{Projection in the $\ell_{\JPsi}$ dimension of the two-dimensional likelihood fit
(in mass and $\ell_{\JPsi}$)
in the bins  $2<\pt<4.5$\GeVc,  $1.2<|y|<1.6$ (left) and
$6.5<\pt<10$\GeVc, $1.6<|y|<2.4$
(right), with their pull distributions (bottom).
}\label{fig:distr_lxy_data}
\end{figure}

The parameter $f_B$ (b fraction) is determined in the same rapidity
regions as used to present the inclusive production cross section but some
\pt bins are grouped, since more events per bin are needed to determine
all fit parameters.
Figure~\ref{fig:distr_lxy_data} shows the projection of the likelihood fits
in two sample bins. The full results are reported
in Table~\ref{tab:lxy_fit_res}, where $f_B$ has been corrected by the
prompt/non-prompt acceptances, as discussed in Section~\ref{sec:Acc_eff}.
The fitting procedure has been tested in five sample bins using toy
experiments, which establish reasonable goodness-of-fit and exclude
the possibility of biases in the $f_B$ determination.

\begin{table}[h!]
\centering
\caption{Fit results for the determination of the fraction of \JPsi mesons from b hadrons in \pt and $|y|$ bins, corrected by the prompt and non-prompt
acceptances. The average \pt~and r.m.s.\ per bin are also quoted. The two uncertainties in the
 b-fraction values are statistical and systematic, respectively.}
\begin{tabular}{ccccc}
\hline
$|y|$ & \pt~ (\GeVc)  & $\langle\pt\rangle$~ (\GeVc) & r.m.s.~ (\GeVc) & b fraction \\
\hline
0--1.2 & $6.5-10.0$           & 8.14 & 0.97 & $0.257 \pm 0.015 \pm 0.014$ \\
& $10.0-30.0$            & 13.50 & 3.53 & $0.395 \pm 0.018 \pm 0.005$ \\
\hline

1.2--1.6 & $2.0-4.5$  & 3.27 & 0.75 & $0.146 \pm 0.021 \pm 0.028$ \\
& $4.5-6.5$          & 5.48 & 0.55 & $0.180 \pm 0.017 \pm 0.019$ \\
& $6.5-10.0$           & 7.89 & 0.93 & $0.203 \pm 0.017 \pm 0.014$ \\
& $10.0-30.0$            & 12.96 & 3.06 & $0.360 \pm 0.031 \pm 0.016$ \\
\hline

1.6--2.4 & $0.00-1.25$ & 0.79 & 0.29 & $0.057 \pm 0.021 \pm 0.042$ \\
& $1.25-2.00$           & 1.60 & 0.21 & $0.087 \pm 0.014 \pm 0.022$ \\
& $2.00-2.75$           & 2.35 & 0.22 & $0.113 \pm 0.013 \pm 0.020$ \\
& $2.75-3.50$         & 3.10 & 0.21 & $0.139 \pm 0.014 \pm 0.010$ \\
& $3.50-4.50$          & 3.96 & 0.29 & $0.160 \pm 0.014 \pm 0.013$ \\
& $4.50-6.50$          & 5.35 & 0.57 & $0.177 \pm 0.012 \pm 0.012$ \\
& $6.50-10.00$           & 7.86 & 0.97 & $0.235 \pm 0.016 \pm 0.012$ \\
& $10.00-30.00$            & 13.11 & 3.23 & $0.374 \pm 0.031 \pm 0.008$ \\
\hline

\end{tabular}
\label{tab:lxy_fit_res}
\end{table}

\begin{figure}[h!]
\centering
\includegraphics[width=8.5cm,angle=90]{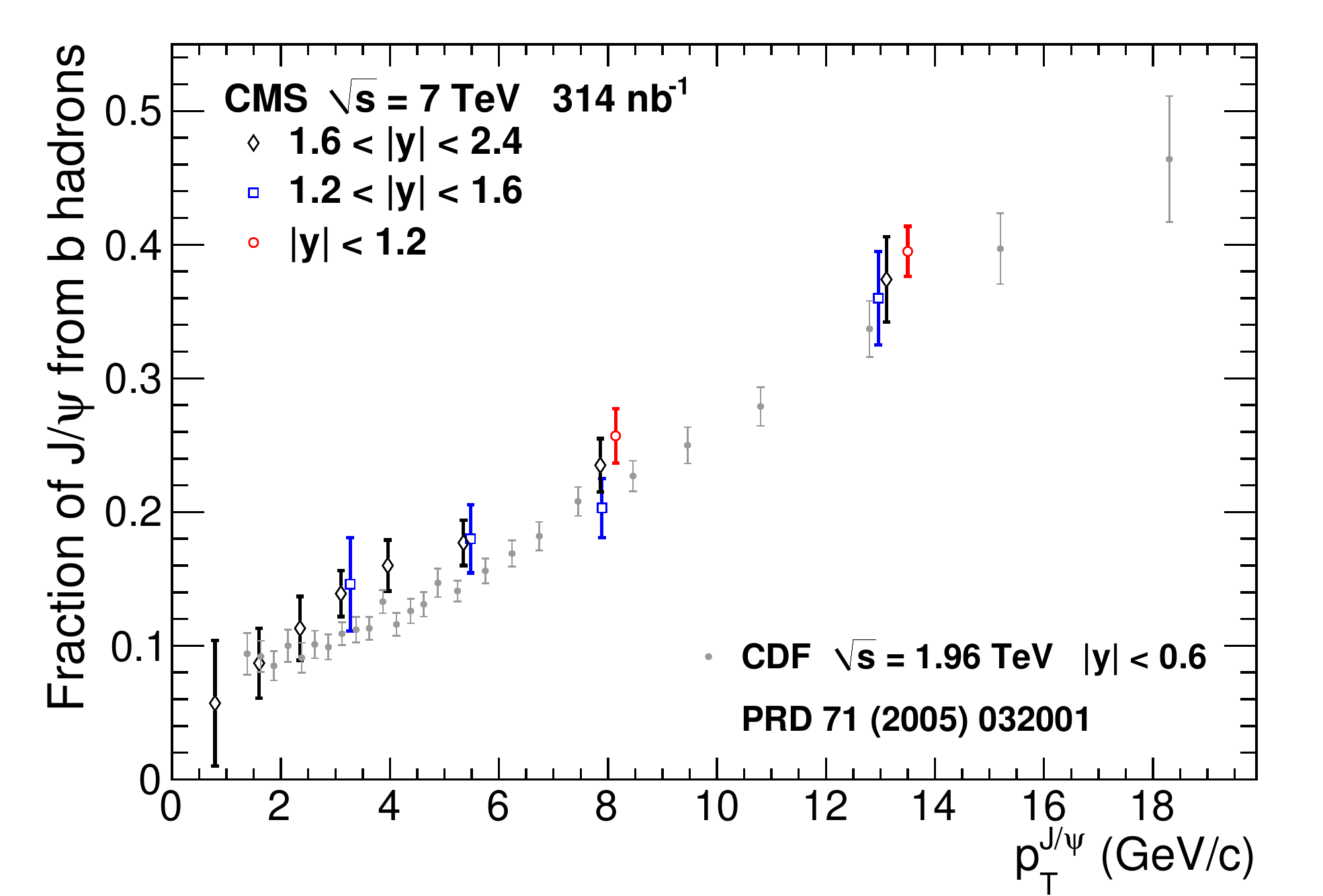}
\caption{Fraction of the \JPsi production cross section originating from b-hadron
decays, as a function of the \JPsi \pt, as measured by CMS in three rapidity bins
and by CDF, at a lower collision energy.}\label{fig:bFraction}
\end{figure}

Figure~\ref{fig:bFraction} shows the measured b fraction. It increases strongly with \pt.
At low \pt, essentially all \JPsi mesons are promptly produced, whereas at
\pt $\sim$\,12\GeVc around one third come from beauty decays. This pattern does not
show a significant change with rapidity (within the current
uncertainties) over the window covered by the CMS
detector.
The CMS results are compared to the higher-precision data of
CDF~\cite{bib-cdfjpsi}, obtained in proton-antiproton collisions at
$\sqrt{s} = 1.96$~TeV. It is interesting to note that the increase with \pt
of the b fraction is very similar between the two experiments, the CMS
points being only slightly higher, despite the different collision energies.

\subsubsection{Systematic uncertainties affecting the b-fraction result}

Several sources of systematic uncertainty have been addressed and are described in the following lines.

\begin{itemize}

\item {\bf Residual misalignments in the tracker.}
The effect of uncertainties in the measured misalignment of the
tracker modules is estimated by reconstructing the data several times
using different sets of alignment constants. These sets reflect the
uncertainty in the constants and, in particular, explore possible
deformations of the tracker which are poorly
constrained by the data~\cite{bib-trackeralignment}.
The largest difference between the
results with the nominal set of constants and with these sets is taken as a
systematic uncertainty.

\item {\bf b-hadron lifetime model}. In an alternative approach, $\ell_{\JPsi}$ is described by a convolution of an exponential
decay with a Gaussian function, which describes the smearing due to the relative
motion of the \JPsi with respect to the parent b hadron.
The difference between the nominal Monte Carlo template model and this alternative is taken as a systematic uncertainty.

\item {\bf Primary vertex estimation}. In an alternative approach, the beam spot as calculated on a run-by-run basis is chosen as the primary vertex in calculating $\ell_{\JPsi}$, and the fit is repeated. The difference is
taken as a systematic uncertainty.

\item {\bf Background}.
The background is fitted using only the sidebands and the result is used as input to the fit in
the signal region. The effect of a $\pm$\,100~MeV/$c^2$ variation in the sideband boundaries is taken as a systematic uncertainty.

\item {\bf $\ell_{\JPsi}$ resolution model}.  The nominal
  (triple-Gaussian) fit model for the decay length resolution is
  compared to a model using two Gaussians only, fixing the
  ``additional'' Gaussian to be zero.  The difference is
  taken as a systematic uncertainty.

\item{\bf Different prompt and non-prompt efficiencies}. The Monte Carlo
simulation predicts small differences between the prompt and
non-prompt \JPsi efficiencies. These are taken into account and the
relative difference assumed as a systematic uncertainty.

\end{itemize}

A summary of all systematic effects and their importance is given in Table~\ref{systab}.

\begin{table}[htb]
\centering \caption{Summary of relative systematic uncertainties in the b-fraction yield (in \%). The variation range over the different \pt bins is given in the
three rapidity regions. In
general, uncertainties are \pt-dependent and decrease with increasing \pt.}
\label{systab}
\begin{tabular}{lccc}
\hline
 & $|y|<1.2$ & $1.2<|y|<1.6$ & $1.6<|y|<2.4$ \\
\hline
Tracker misalignment & $0.5-0.7$ & $0.9-4.6$ & $0.7-9.1$ \\
b-lifetime model & $0.0-0.1$ & $0.5-4.8$ & $0.5-11.2$ \\
Vertex estimation & $0.3$ & $1.0-12.3$ & $0.9-65.8$ \\
Background fit &  $0.1-4.7$ & $0.5-9.5$ & $0.2-14.8$ \\
Resolution model & $0.8-2.8$ & $1.3-13.0$ & $0.4-30.2$ \\
Efficiency & $0.1-1.1$ & $0.3-1.3$ & $0.2-2.4$ \\
\hline
\end{tabular}
\end{table}

\subsubsection{Prompt and non-prompt \JPsi production cross sections}

The prompt
\JPsi~cross section and the cross section from
b-hadron decays, together with their statistical and
systematic uncertainties, are given in
Tables~\ref{tab:prnonpr} and~\ref{tab:prnonpr2}, respectively,
for the different polarization scenarios considered in Section~\ref{inclusive}.

%\begin{sidewaystable}[h!]
\begin{table}[h!]
\caption{Differential prompt \JPsi cross sections for each polarization scenario
considered: unpolarized ($\lambda_\theta=0$), full longitudinal polarization
($\lambda_\theta =-1$) and full transverse polarization ($\lambda_\theta=+1$)
in the Collins-Soper (CS) or the Helicity (HX) frames~\cite{bib-faccioli}.
For the unpolarized case, the first error is statistical and the second is systematic; for the others the total error is given.}
\label{tab:prnonpr}
\begin{center}
{\small
\begin{tabular}{lccccc}\hline
 \pt & \multicolumn{5}{c}{$BR(\JPsi\to\mu^+\mu^-)\cdot\frac{d^2\sigma_{\mathrm{prompt}}}{d\pt dy}$  (nb/\GeVc)} \\
  (\GeVc)         & $\lambda_\theta=0$  & $\lambda_\theta^{CS} =-1$ & $\lambda_\theta^{CS} =+1$ & $\lambda_\theta^{HX} =-1$& $\lambda_\theta^{HX} =+1$ \\
\hline	
\multicolumn{6}{c}{$|y|<1.2$}\\
\hline	
$6.5-10.0$     &  $3.76 \pm 0.13 \pm 0.47$ & $4.63 \pm 0.60$ & $3.45 \pm 0.45$ & $2.63 \pm 0.34$ & $4.79 \pm 0.62$ \\
$10.0-30.0$    & $0.134 \pm 0.033 \pm 0.016$ & $0.161 \pm 0.044$ & $0.123 \pm 0.033$ & $0.099 \pm 0.026$ & $0.164 \pm 0.045$ \\
\hline	
\multicolumn  {6}{c}{$1.2<|y|<1.6$}\\
\hline	
$2.0-4.5$      &  $50.6 \pm 3.6 \pm 8.4$ & $36.4 \pm 6.5$ & $63.6 \pm 11.6$ & $36.3 \pm 6.5$ & $63.1 \pm 11.4$ \\
$4.5-6.5$      &  $18.4 \pm 0.7 \pm 2.4$ & $17.3 \pm 2.3$ & $19.1 \pm 2.6$ & $13.3 \pm 1.8$ & $22.7 \pm 3.1$ \\
$6.5-10.0$     &  $3.85 \pm 0.15 \pm 0.44$ & $4.11 \pm 0.49$ & $3.74 \pm 0.45$ & $2.87 \pm 0.34$ & $4.67 \pm 0.56$ \\
$10.0-30.0$    &  $0.116 \pm 0.009 \pm 0.014$ & $0.127 \pm 0.018$ & $0.111 \pm 0.015$ & $0.093 \pm 0.013$ & $0.133 \pm 0.019$ \\
\hline	
\multicolumn  {6}{c}{$1.6<|y|<2.4$}\\
\hline	
$0.00-1.25$    &  $71.9 \pm 2.4 \pm 11.2$ & $49.7 \pm 7.9$ & $92.5 \pm 14.7$ & $51.0 \pm 8.1$ & $90.3 \pm 14.3$ \\
$1.25-2.00$    &  $116.2 \pm 3.5 \pm 16.8$ & $80.8 \pm 11.9$ & $149.1 \pm 22.0$ & $86.7 \pm 12.8$ & $140.7 \pm 20.8$ \\
$2.00-2.75$    &  $93.7 \pm 3.4 \pm 12.4$ & $65.8 \pm 9.1$ & $118.8 \pm 16.3$ & $72.7 \pm 10.0$ & $110.3 \pm 15.2$ \\
$2.75-3.50$    &  $62.6 \pm 2.0 \pm 7.9$ & $44.5 \pm 5.7$ & $78.8 \pm 10.2$ & $49.1 \pm 6.4$ & $72.7 \pm 9.5$ \\
$3.50-4.50$    &  $37.4 \pm 1.1 \pm 4.9$ & $27.4 \pm 3.7$ & $45.7 \pm 6.2$ & $29.9 \pm 4.1$ & $42.8 \pm 5.8$ \\
$4.50-6.50$    &  $15.2 \pm 0.4 \pm 2.0$ & $11.9 \pm 1.6$ & $18.0 \pm 2.4$ & $12.6 \pm 1.7$ & $17.1 \pm 2.3$ \\
$6.50-10.00$   &  $3.08 \pm 0.11 \pm 0.37$ & $2.79 \pm 0.35$ & $3.36 \pm 0.42$ & $2.64 \pm 0.33$ & $3.37 \pm 0.42$ \\
$10.00-30.00$  & $0.093 \pm 0.007 \pm 0.012$ & $0.092 \pm 0.014$ & $0.096 \pm 0.014$ & $0.082 \pm 0.012$ & $0.100 \pm 0.015$ \\
\hline
\end{tabular}
}
\end{center}
\end{table}
\begin{table}[htbp!]
\caption{Differential non-prompt \JPsi~cross section times the \JPsi branching ratio to dimuons, assuming the polarization measured by the BaBar
experiment~\cite{babarpol} at the $\Upsilon$(4S).
The first uncertainty is statistical and the second is systematic.}
\label{tab:prnonpr2}
\begin{center}
\begin{tabular}{lc}\hline
\ptjpsi~ & $BR(\JPsi\to\mu^+\mu^-)\cdot\frac{d^2\sigma_{\mathrm{non-prompt}}}{d\pt dy}$   \\
 (\GeVc)  &  (nb/\GeVc) \\
\hline	
\multicolumn{2}{c}{$|y|<1.2$}\\
\hline	
$6.5-10.0$     & $1.30 \pm 0.08 \pm 0.19$ \\
$10.0-30.0$    & $0.087 \pm 0.024 \pm 0.010$\\
\hline	
\multicolumn{2}{c}{$1.2<|y|<1.6$}\\
\hline	
$2.0-4.5$      & $8.67 \pm 1.36 \pm 2.71$ \\
$4.5-6.5$      & $4.04 \pm 0.41 \pm 0.79$ \\
$6.5-10.0$     & $0.98 \pm 0.09 \pm 0.11$ \\
$10.0-30.0$    & $0.065 \pm 0.007 \pm 0.008$ \\
\hline	
\multicolumn{2}{c}{$1.6<|y|<2.4$}\\
\hline	
$0.00-1.25$    & $4.31 \pm 1.59 \pm 3.54$ \\
$1.25-2.00$    & $11.0 \pm 1.8 \pm 4.2$ \\
$2.00-2.75$    & $11.9 \pm 1.4 \pm 3.4$ \\
$2.75-3.50$    & $10.1 \pm 1.1 \pm 1.6$ \\
$3.50-4.50$    & $7.19 \pm 0.65 \pm 1.25$ \\
$4.50-6.50$    & $3.28 \pm 0.24 \pm 0.53$ \\
$6.50-10.00$   & $0.95 \pm 0.07 \pm 0.13$\\
$10.00-30.00$  & $0.055 \pm 0.005 \pm 0.007$\\
\hline
\end{tabular}
\end{center}
\end{table}

The total cross section for prompt \JPsi~production times $BR(\JPsi\to\mu^+\mu^-)$, for the unpolarized production scenario,
has been obtained by integrating the differential cross section over \pt between 6.5 and 30\GeVc
and over rapidity between $-2.4$ and $2.4$,
\begin{equation}
BR(\JPsi\to\mu^+\mu^-)\cdot\sigma(\mathrm{pp}\rightarrow \mathrm{prompt}~\JPsi) = 70.9 \pm 2.1 \pm 3.0 \pm 7.8~\mathrm{nb}\quad,
\end{equation}
where the three uncertainties are statistical, systematic and due to the measurement of the integrated luminosity, respectively.
Similarly, the cross section of non-prompt \JPsi mesons from b-hadron decays, times $BR(\JPsi\to\mu^+\mu^-)$, is
\begin{equation}
BR(\JPsi\to\mu^+\mu^-)\cdot\sigma(\mathrm{pp} \rightarrow b X\rightarrow \JPsi X) = 26.0 \pm 1.4 \pm 1.6 \pm 2.9~\mathrm{nb}\quad .
\end{equation}
The sum of these two cross sections differs slightly from the inclusive value, which was determined
assuming a b fraction taken from Monte Carlo expectations.

%% file: Comparison.tex
\section{Comparison with theoretical calculations}\label{sec:comparison}

\begin{figure}[h!]
\centering
\includegraphics[angle=90,width=6cm]{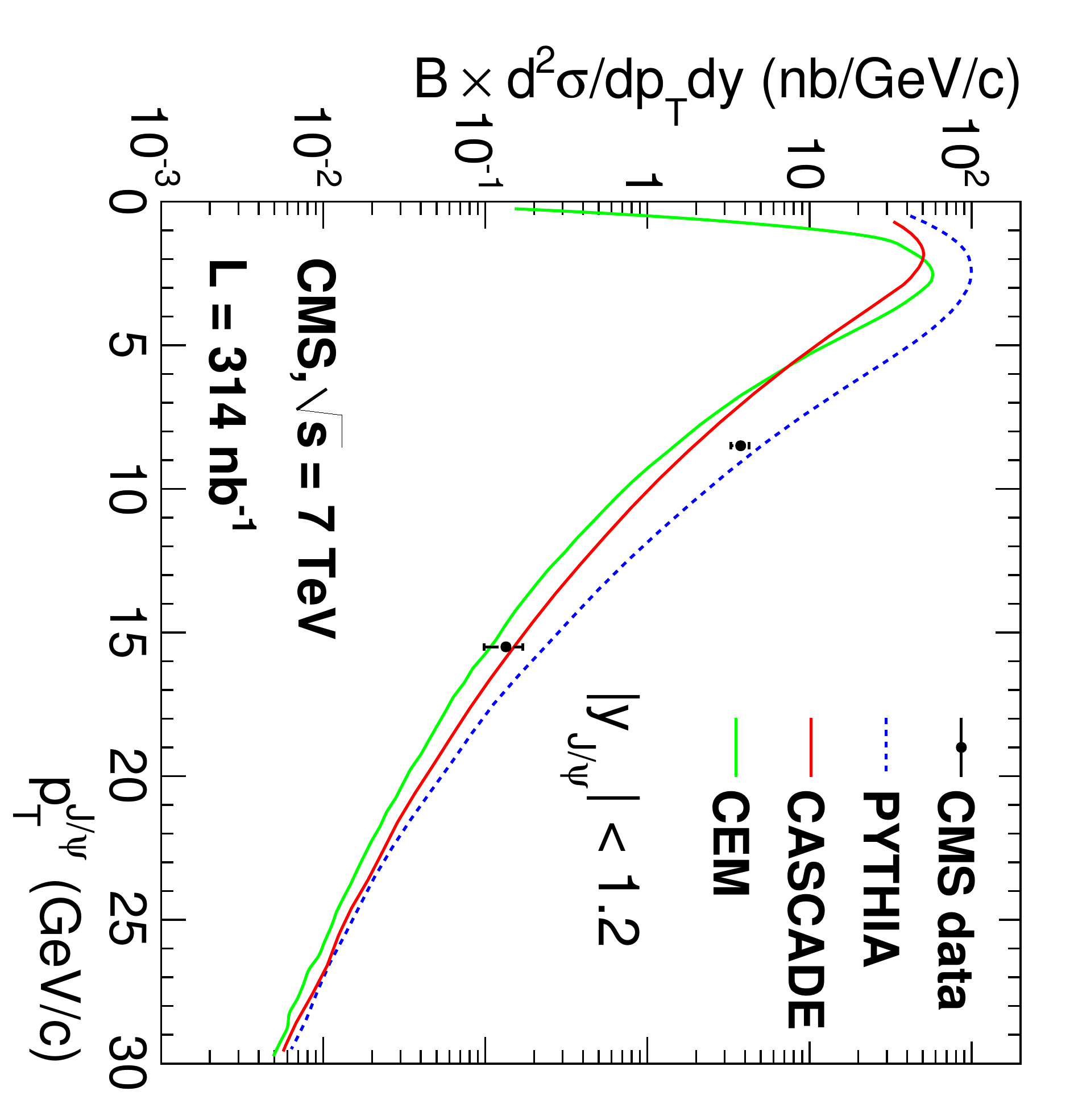}
\includegraphics[angle=90,width=6cm]{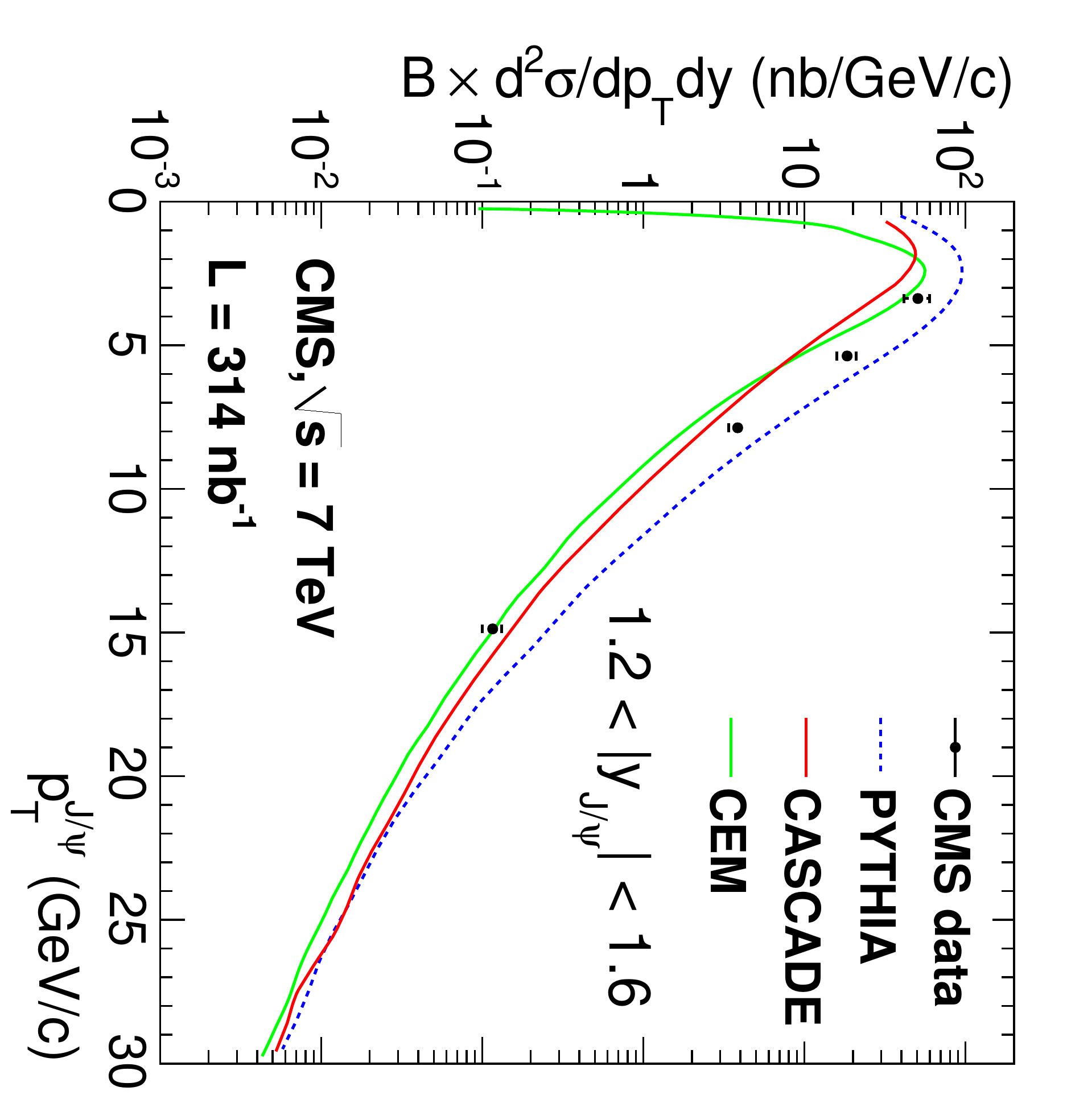} \\
\includegraphics[angle=90,width=6cm]{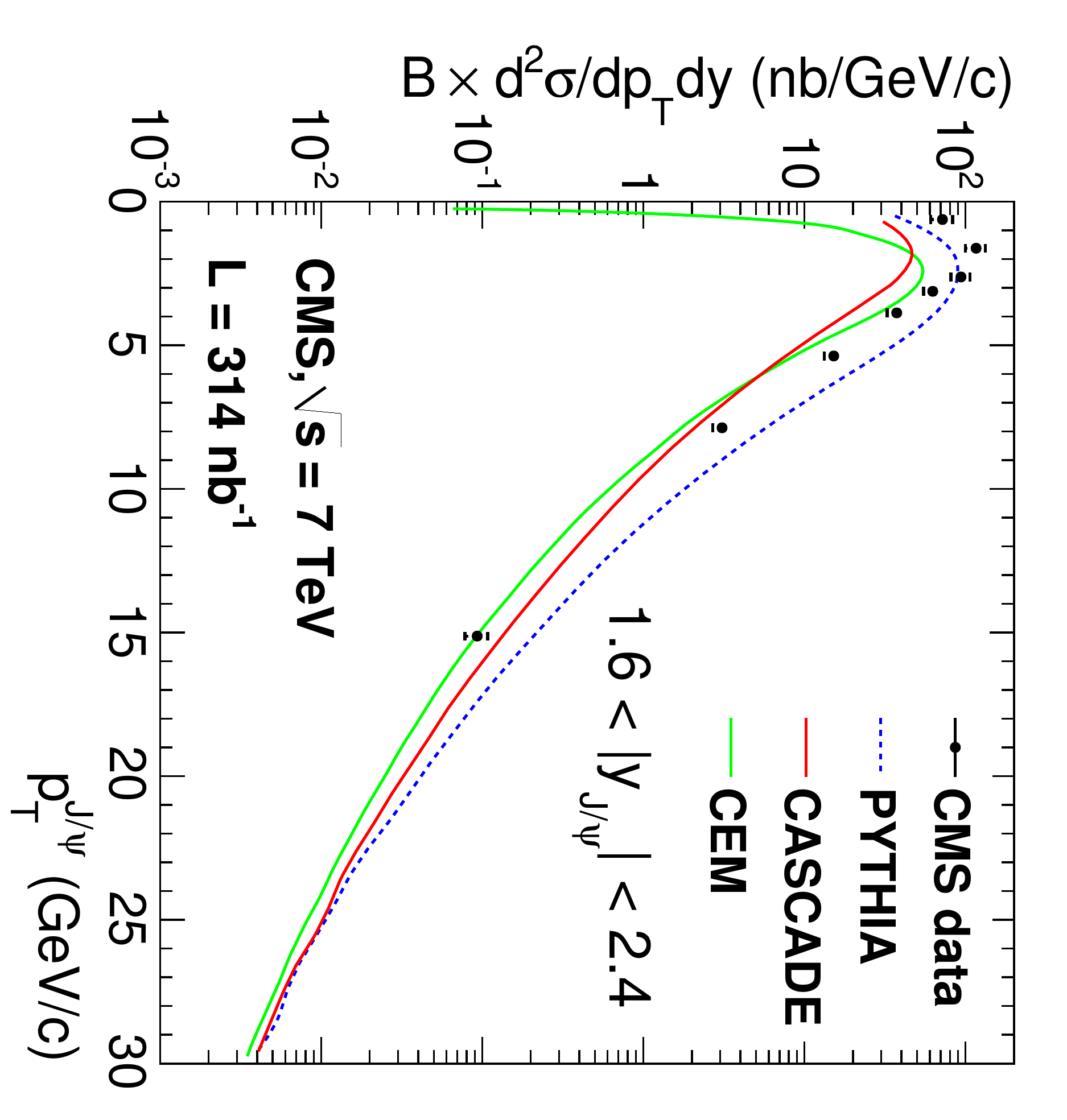}
\caption{ \small{Differential prompt \JPsi production cross section, as a
function of \pt~for the three different rapidity intervals.
The data points are compared with three different models, using the PYTHIA curve to
calculate the abscissa where they are plotted~\cite{wyatt}.
}}
\label{fig:theoryprompt}
\end{figure}

The prompt \JPsi differential production cross sections, in
the rapidity ranges considered in the analysis,
as summarized in Table~\ref{tab:prnonpr},
were compared with
calculations made with the Pythia~\cite{bib-PYTHIA} and
CASCADE~\cite{bib-CASCADE,Jung:2010si} event generators, as well as with the Color
Evaporation Model (CEM)~\cite{ColorEvaporationModel,Halzen:1977rs,Fritzsch:1977ay,Gluck:1977zm,Barger:1979js}.
These calculations include the contributions to the prompt \JPsi yield
due to feed-down decays from heavier charmonium states ($\chi_c$ and
$\psi(2S)$) and can, therefore, be directly compared to the measured
data points, as shown in Fig.~\ref{fig:theoryprompt}. In contrast, it is
not
possible to compare our measurement with the predictions of models such as the Color-Singlet Model (including higher-order corrections)~\cite{bib-campbell, Artoisenet:2008fc, Artoisenet:2007xi,  Lansberg:2008gk}
or
the LO NRQCD model (which includes singlet and octet components)~\cite{bib-octet1,bib-octet2}, because they are only available for the \emph{direct} \JPsi production component, while the measurements include a
significant
contribution from feed-down decays,
of the order of 30\%~\cite{Ma:2010vd, Faccioli:2008ir}.
At forward rapidity and low \pt the calculations underestimate the measured yield.

\begin{figure}[h!]
\centering
\includegraphics[angle=90,width=6.5cm]{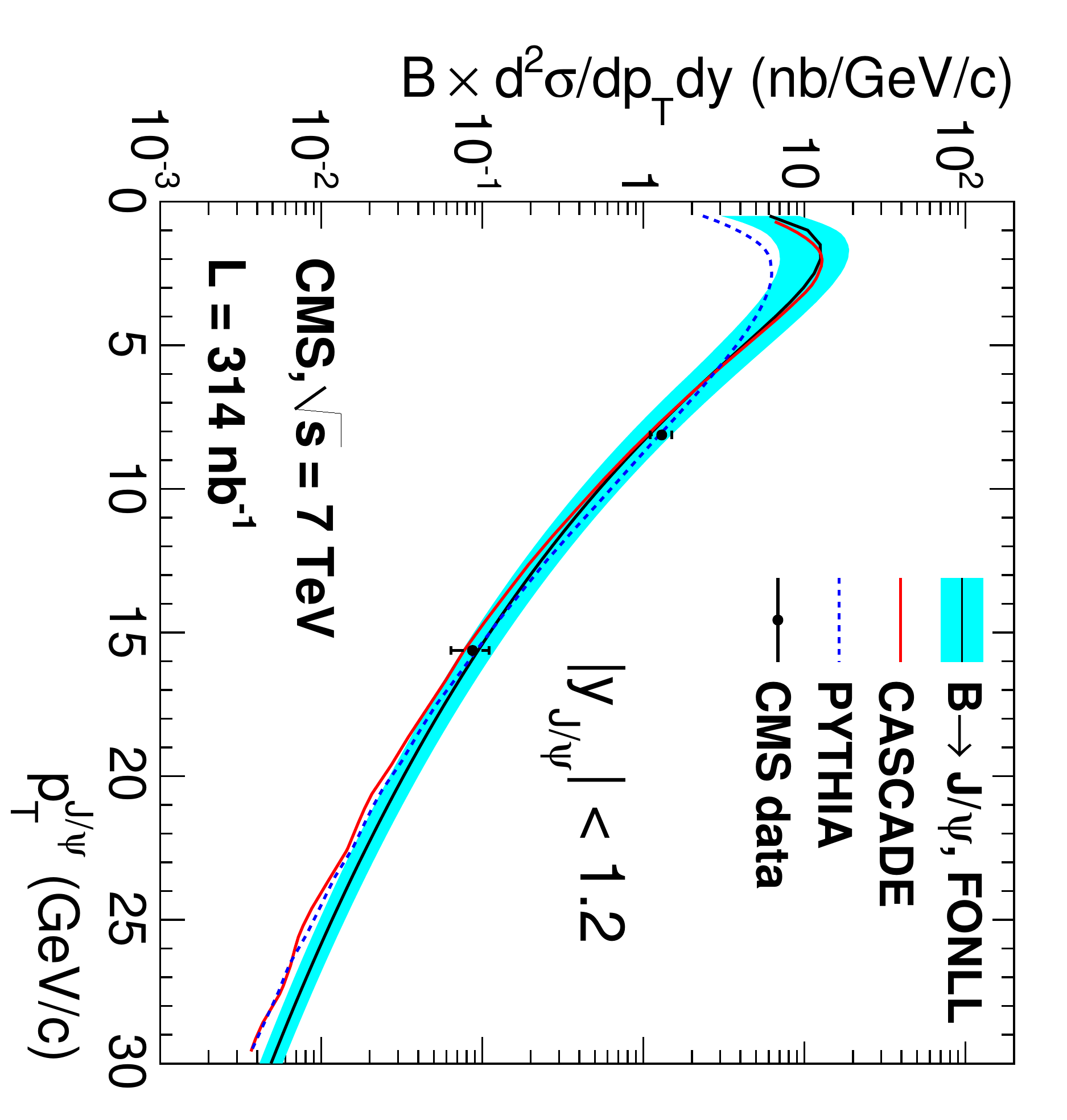}
\includegraphics[angle=90,width=6.5cm]{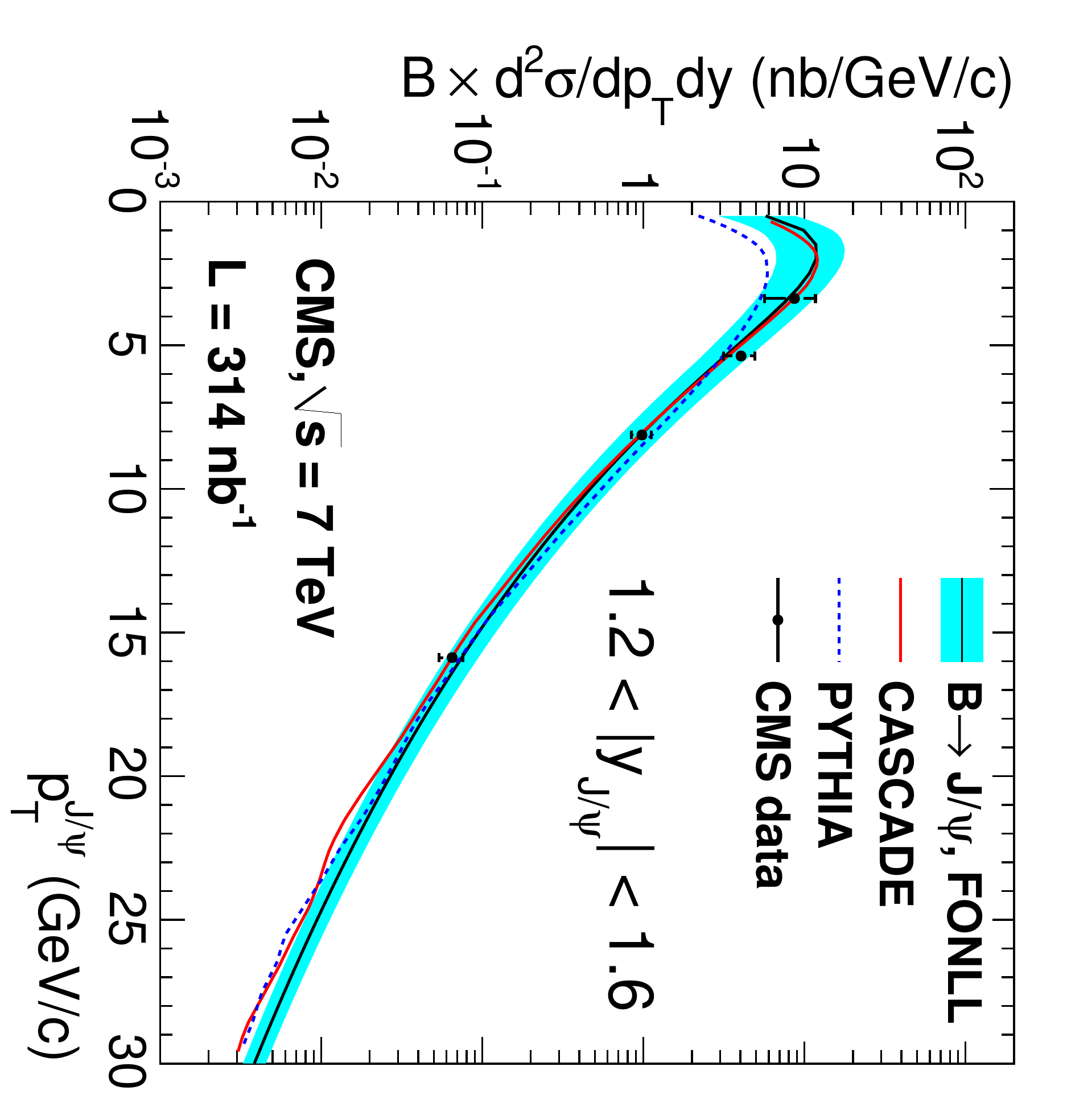} \\
\includegraphics[angle=90,width=6.5cm]{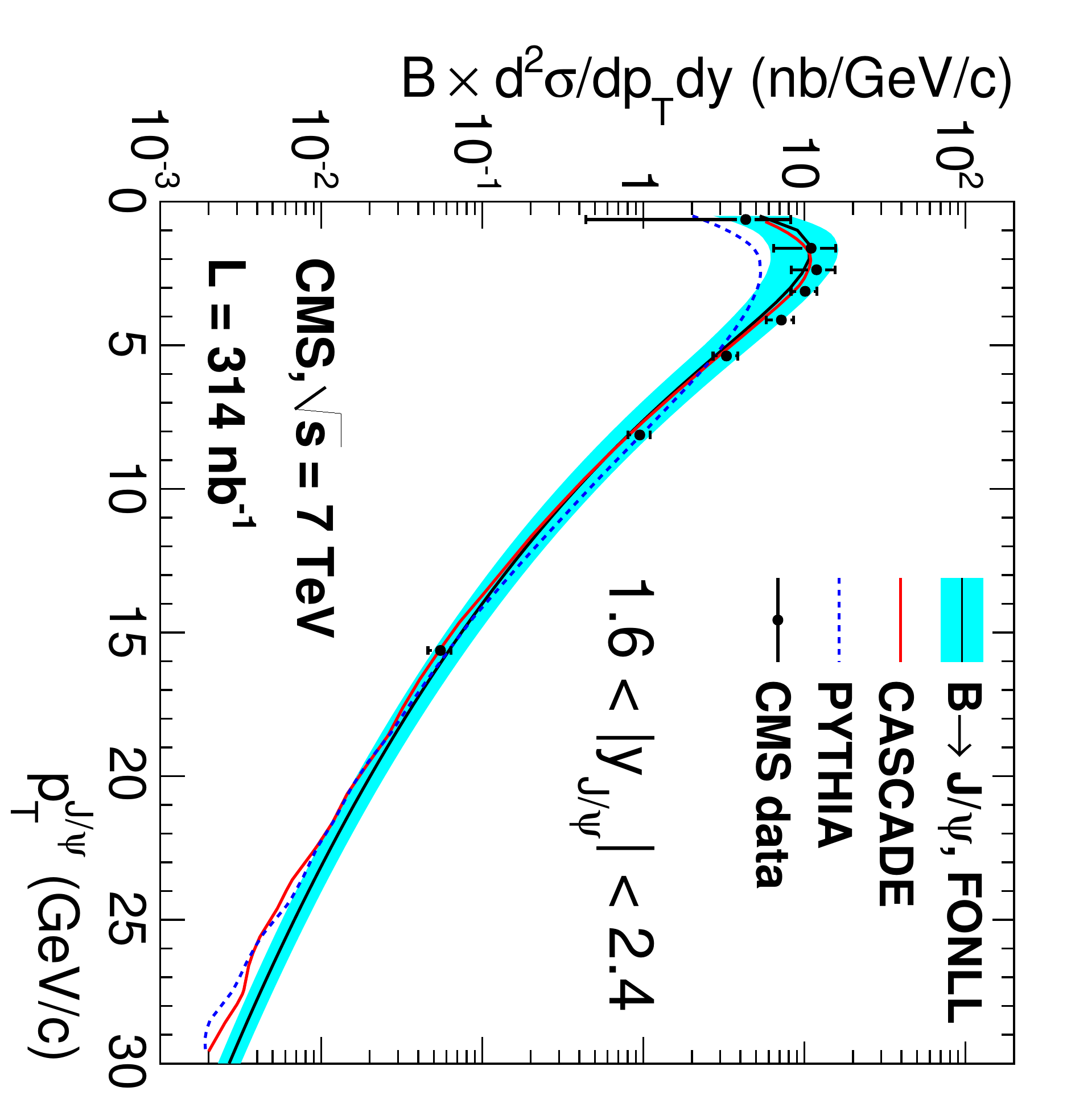}
\caption{
\small{Differential non-prompt \JPsi production cross section, as a
function of \pt~for the three different rapidity intervals.
The data points are compared with three different models, using the PYTHIA curve to
calculate the abscissa where they are plotted~\cite{wyatt}.
}
}
\label{fig:theorynonprompt}
\end{figure}

The non-prompt \JPsi differential production cross sections,
as summarized in Table~\ref{tab:prnonpr2}, have been compared with calculations
made with the Pythia and CASCADE Monte Carlo generators, and in the FONLL
framework~\cite{cacciar2}. The measured results are presented in
Fig.~\ref{fig:theorynonprompt} and show a good agreement with the calculations.

%% file: Conclusions.tex
\section{Conclusions}\label{sec:conclusions}

We have presented the first measurement of the
\JPsi~production cross section
in pp collisions at $\sqrt{s}=7$~TeV, based on
314~nb$^{-1}$ of integrated luminosity
  collected by the CMS experiment
 during the first months of LHC operation.

The \pt differential \JPsi production cross section,  in the dimuon decay channel,
has been measured in three rapidity ranges, starting
at zero \pt for $1.6<|y|<2.4$,
at 2~\GeVc for $1.2<|y|<1.6$, and
at 6.5~\GeVc for $|y|<1.2$.
The measured total cross section for prompt \JPsi production in the unpolarized scenario,
in the dimuon decay channel, is
%\begin{equation}
%\nonumber
$$
\sigma(pp\rightarrow\JPsi +X)\cdot\mbox{BR}(\JPsi\to\mu^+\mu^-) = 70.9 \pm 2.1\mbox{(stat)} \pm 3.0\mbox{(syst)} \pm 7.8\mbox{(luminosity)} ~\mathrm{nb}\quad ,
$$
for transverse momenta between $6.5$ and $30$\GeVc and in the rapidity range  $|y|<2.4$. Aside from the luminosity contribution, the systematic uncertainty is dominated by the statistical precision of the muon efficiency determination
from data.

The measured total cross section times $BR(\JPsi\to\mu^+\mu^-)$ for \JPsi
production due to b-hadron decays, for $6.5<\pt<30$\GeVc and $|y|<2.4$, is
$$
\sigma(\mathrm{pp}\rightarrow b X\rightarrow \JPsi X)\cdot \mbox{BR}(\JPsi\to\mu^+\mu^-) = 26.0 \pm 1.4~\mbox{(stat)} \pm 1.6~\mbox{(syst)} \pm 2.9~\mbox{(luminosity)}~\mathrm{nb}\quad .  $$

The differential prompt and non-prompt measurements have been compared
with theoretical calculations. A reasonable agreement is found between data and theory
for the non-prompt case while the measured prompt \JPsi cross section exceeds the
expectations at forward rapidity and low \pt.

{\bf Acknowledgments}\\
We would like to thank Pierre Artoisenet, Jean-Philippe Lansberg, and Ramona Vogt for providing their theoretical predictions in the prompt production models
and Matteo Cacciari for predictions in the FONLL scheme.

We wish to congratulate our colleagues in the CERN accelerator departments for the excellent performance of the LHC machine. We thank the technical and administrative staff at CERN and other CMS institutes. This work was supported by the Austrian Federal Ministry of Science and Research; the Belgium Fonds de la Recherche Scientifique, and Fonds voor Wetenschappelijk Onderzoek; the Brazilian Funding Agencies (CNPq, CAPES, FAPERJ, and FAPESP); the Bulgarian Ministry of Education and Science; CERN; the Chinese Academy of Sciences, Ministry of Science and Technology, and National Natural Science Foundation of China; the Colombian Funding Agency (COLCIENCIAS); the Croatian Ministry of Science, Education and Sport; the Research Promotion Foundation, Cyprus; the Estonian Academy of Sciences and NICPB; the Academy of Finland, Finnish Ministry of Education, and Helsinki Institute of Physics; the Institut National de Physique Nucl\'eaire et de Physique des Particules~/~CNRS, and Commissariat \`a l'\'Energie Atomique, France; the Bundesministerium f\"ur Bildung und Forschung, Deutsche Forschungsgemeinschaft, and Helmholtz-Gemeinschaft Deutscher Forschungszentren, Germany; the General Secretariat for Research and Technology, Greece; the National Scientific Research Foundation, and National Office for Research and Technology, Hungary; the Department of Atomic Energy, and Department of Science and Technology, India; the Institute for Studies in Theoretical Physics and Mathematics, Iran; the Science Foundation, Ireland; the Istituto Nazionale di Fisica Nucleare, Italy; the Korean Ministry of Education, Science and Technology and the World Class University program of NRF, Korea; the Lithuanian Academy of Sciences; the Mexican Funding Agencies (CINVESTAV, CONACYT, SEP, and UASLP-FAI); the Pakistan Atomic Energy Commission; the State Commission for Scientific Research, Poland; the Funda\c{c}\~ao para a Ci\^encia e a Tecnologia, Portugal; JINR (Armenia, Belarus, Georgia, Ukraine, Uzbekistan); the Ministry of Science and Technologies of the Russian Federation, and Russian Ministry of Atomic Energy; the Ministry of Science and Technological Development of Serbia; the Ministerio de Ciencia e Innovaci\'on, and Programa Consolider-Ingenio 2010, Spain; the Swiss Funding Agencies (ETH Board, ETH Zurich, PSI, SNF, UniZH, Canton Zurich, and SER); the National Science Council, Taipei; the Scientific and Technical Research Council of Turkey, and Turkish Atomic Energy Authority; the Science and Technology Facilities Council, UK; the US Department of Energy, and the US National Science Foundation.
Individuals have received support from the Marie-Curie IEF program (European Union); the Leventis Foundation; the A. P. Sloan Foundation; the Alexander von Humboldt Foundation; the Associazione per lo Sviluppo Scientifico e Tecnologico del Piemonte (Italy); the Belgian Federal Science Policy Office; the Fonds pour la Formation \`a la Recherche dans l'\'industrie et dans l'\'Agriculture (FRIA-Belgium); and the Agentschap voor Innovatie door Wetenschap en Technologie (IWT-Belgium).

%% We congratulate and express our gratitude to our colleagues in the CERN
%% accelerator departments for the excellent performance of the LHC. We thank the
%% technical and administrative staff at CERN and other CMS Institutes, and
%% acknowledge support from: FMSR (Austria); FNRS and FWO (Belgium); CNPq, CAPES,
%% FAPERJ, and FAPESP (Brazil); MES (Bulgaria); CERN; CAS, MoST, and NSFC (China);
%% COLCIENCIAS (Colombia); MSES (Croatia); RPF (Cyprus); Academy of Sciences and
%% NICPB (Estonia); Academy of Finland, ME, and HIP (Finland); CEA and
%% CNRS/IN2P3 (France); BMBF, DFG, and HGF (Germany); GSRT (Greece); OTKA and
%% NKTH (Hungary); DAE and DST (India); IPM (Iran); SFI (Ireland); INFN (Italy);
%% NRF (Korea); LAS (Lithuania); CINVESTAV, CONACYT, SEP, and UASLP-FAI
%% (Mexico); PAEC (Pakistan); SCSR (Poland); FCT (Portugal);
%% JINR (Armenia, Belarus, Georgia, Ukraine, Uzbekistan); MST and MAE (Russia);
%% MSTDS (Serbia); MICINN and CPAN (Spain); Swiss Funding Agencies
%% (Switzerland); NSC (Taipei); TUBITAK and TAEK (Turkey); STFC (United Kingdom);
%% DOE and NSF (USA). Individuals have received support from the Marie-Curie
%% IEF program (European Union); the Leventis Foundation; the A. P. Sloan
%% Foundation; and the Alexander von Humboldt Foundation.

%% file: BPH-10-002-authorlist.tex
\textbf{Yerevan Physics Institute,  Yerevan,  Armenia}\\*[0pt]
V.~Khachatryan, A.M.~Sirunyan, A.~Tumasyan
\vskip\cmsinstskip
\textbf{Institut f\"{u}r Hochenergiephysik der OeAW,  Wien,  Austria}\\*[0pt]
W.~Adam, T.~Bergauer, M.~Dragicevic, J.~Er\"{o}, C.~Fabjan, M.~Friedl, R.~Fr\"{u}hwirth, V.M.~Ghete, J.~Hammer\cmsAuthorMark{1}, S.~H\"{a}nsel, C.~Hartl, M.~Hoch, N.~H\"{o}rmann, J.~Hrubec, M.~Jeitler, G.~Kasieczka, W.~Kiesenhofer, M.~Krammer, D.~Liko, I.~Mikulec, M.~Pernicka, H.~Rohringer, R.~Sch\"{o}fbeck, J.~Strauss, A.~Taurok, F.~Teischinger, W.~Waltenberger, G.~Walzel, E.~Widl, C.-E.~Wulz
% ...
\vskip\cmsinstskip
% ...
1:~~Also at CERN, European Organization for Nuclear Research, Geneva, Switzerland\\
% ...
